\documentclass[]{aastex631}

\usepackage{multirow}
\usepackage{amsmath}

\received{June 9, 2023}
\accepted{August 28, 2023}

\submitjournal{APJ}

\begin{document}

\title{Probabilistic solar flare forecasting using historical magnetogram data}

\correspondingauthor{Kiera van der Sande}
\email{kiera.vandersande@swri.org}

\author[0000-0001-6606-6481]{Kiera van der Sande}
\affiliation{Southwest Research Institute \\
1050 Walnut St., Suite 300, \\
Boulder, CO 80302, USA}

\author[0000-0002-4716-0840]{Andr\'es Mu\~noz Jaramillo}
\affiliation{Southwest Research Institute \\
1050 Walnut St., Suite 300, \\
Boulder, CO 80302, USA}

\author{Subhamoy Chatterjee}
\affiliation{Southwest Research Institute \\
1050 Walnut St., Suite 300, \\
Boulder, CO 80302, USA}



\begin{abstract}

Solar flare forecasting research using machine learning (ML) has focused on high resolution magnetogram data from the SDO/HMI era covering Solar Cycle 24 and the start of Solar Cycle 25, with some efforts looking back to SOHO/MDI for data from Solar Cycle 23. In this paper, we consider over 4 solar cycles of daily historical magnetogram data from multiple instruments. This is the first attempt to take advantage of this historical data for ML-based flare forecasting. We apply a convolutional neural network (CNN) to extract features from full-disk magnetograms together with a logistic regression model to incorporate scalar features based on magnetograms and flaring history. We use an ensemble approach to generate calibrated probabilistic forecasts of M-class or larger flares in the next 24 hours. Overall, we find that including historical data improves forecasting skill and reliability. We show that single frame magnetograms do not contain significantly more relevant information than can be summarized in a small number of scalar features, and that flaring history has greater predictive power than our CNN-extracted features. This indicates the importance of including temporal information in flare forecasting models.

\end{abstract}

\keywords{Solar flares, Convolutional neural networks, Space weather, Solar Activity, Solar magnetic fields}


\section{Introduction} \label{sec:intro}


    
Solar flares are outbursts of electromagnetic radiation associated with intense releases of magnetic energy {triggered by magnetic reconnection in} active regions on the Sun's surface. They can interfere with radio, radar and Global Navigation Satellite System (GNSS) signals. As the first indication of {magnetic reconnection events} that we receive on Earth, solar flares {can} also serve as precursors to additional space weather phenomena including coronal mass ejections (CMEs) and solar energetic particle (SEP) events. With advance warning, the effects of these events can be mitigated to some extent. Because of this, increasing {the} accuracy and reliability {of flare forecasts} is a key goal of space weather forecasting research. 
Current operational flare forecasting systems rely on sunspot classification tables \citep{McIntosh_1990} and human forecasters-in-the-loop. However, there has been a large effort to develop statistical and machine learning (ML) techniques to improve solar flare forecasts. 

Much of the ML-based flare forecasting research has used scalar features associated with solar activity, for example, the Spaceweather HMI Active Region Patch (SHARP) parameters \citep{bobra:2014}. Models trained on the SHARPs parameters or similar physics-based features include linear discriminant analysis \citep{Leka2007}, logistic regression \citep{Yuan2010}, random forests \citep{Florios2018,Nishizuka2017}, and support vector machines \citep{Bobra2015,Florios2018}, as well as deep-learning models such as multilayer perceptrons \citep{Florios2018,Nishizuka2018} and long short-term memories \citep{Chen2019}. There has been an effort to compute topology-based features from magnetograms, and these have also been shown to be predictive of flaring \citep{Deshmukh2020,Sun2021}. Other recent works including \citet{deshmukh2022decreasing,Guastavino2022,Huang2018,Li2020,Sun2022} have instead explored use of convolutional neural networks (CNNs) for automatically extracting flaring-related features from magnetograms. 

One challenge of forecasting solar flares is that large flares occur infrequently. Solar flares are measured by their peak X-ray flux emission and classified as A, B, C, M and X flares with M-flares having a peak intensity greater than $10^{-5}  W/m^2$ and X-flares greater than $10^{-6} W/m^2$ {in the 0.1-0.8 nm wavelength band}. M- and X-class flares are the events large enough to impact human activities, hence forecasters tend to issue 24, 48 and 72 hour forecasts of flares above the M-class threshold. This rare event problem is exacerbated by the fact that modern high resolution instruments have been operating during a relatively quiet solar cycle. Since the launch of the Solar Dynamics Observatory (SDO) with the Helioseismic and Magnetic Imager (HMI) in 2010, there have been fewer than 750 M/X flares. An additional issue is that current flare catalogs are inconsistent and may contain errors, as has been noted in \citet{Angryk2020, vanderSande2022}. This results in highly imbalanced data for training ML models in an operational setting. Obtaining more flaring events is thus an area of opportunity. There has been some effort to extend ML models over an additional solar cycle by incorporating data from the Michelson Doppler Imager (MDI) onboard the Solar and Heliospheric Observatory (SOHO). In \citet{Sun2022} including both MDI and HMI datasets was found to be beneficial. However, in \citet{Leka2019b}, no benefit was found when comparing models trained on both datasets.

\begin{figure}[hbtp]
    \centering
    \includegraphics[width=\textwidth]{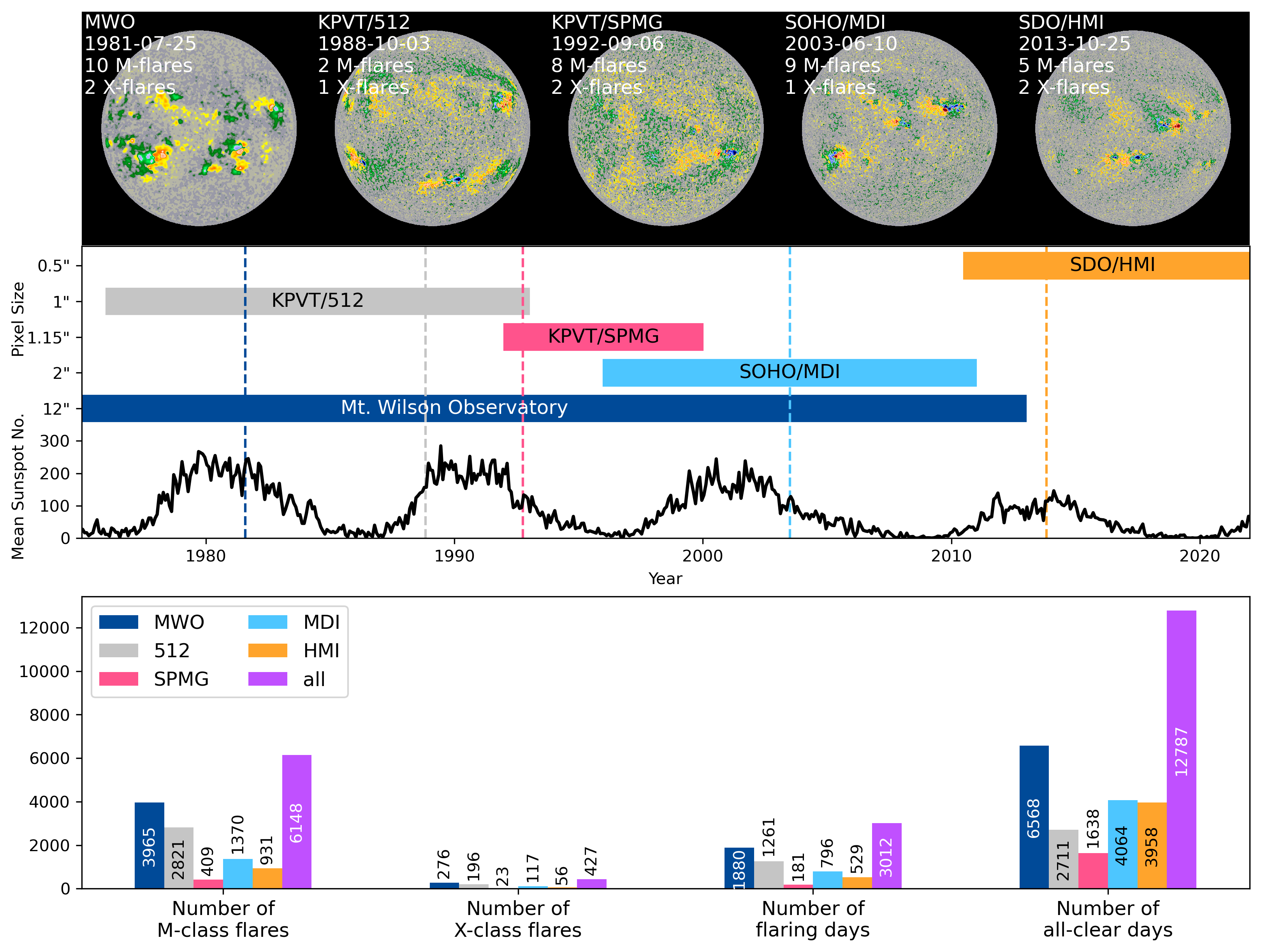}
    \caption{Duration and overlap of historical magnetogram instruments, and the corresponding flares captured by them. {Top, sample magnetograms from each instrument on the given date, as well as the number of M or X flares recorded within a 24 hour window.}}
    \label{fig:flareshistorical}
\end{figure}

Here, we consider extending our dataset to include historical line-of-sight (LOS) magnetograms spanning over 40 years of observations and 4 solar cycles. This is the first inclusion of this historical data for ML-based flare forecasting. Including multiple solar cycles greatly increases the number of flares in the dataset -- there are nearly 10 times as many M/X flares over this period compared to the HMI era alone. Fig. \ref{fig:flareshistorical} illustrates the duration and overlap of the instruments we consider: Mount Wilson Observatory (MWO) {\citep{howard1983mount,ulrich2002mount}}, Kitt Peak Vacuum Telescope (KPVT) 512 and SPMG {\citep{livingston1976kitt}}, SOHO/MDI {\citep{scherrer1995solar}}, and SDO/HMI {\citep{scherrer2012helioseismic}}. We note that the resolution and quality across these instruments varies widely. Despite the increase in temporal coverage of the combined instruments, older data is only available at a daily cadence compared to the 12 minute cadence now available with HMI. We seek to answer the question of whether including historical data can improve ML-based forecasting models. We do so using ML models which take single frame full-disk magnetograms as input, as well as scalar features derived from the magnetograms and flaring history, and output a probability of flaring in the next 24 hours. 

We focus on probabilistic flare forecasting using an ensemble approach. Incorporating probability rather than just a binary forecast offers a more nuanced picture, which is beneficial for stakeholders with varying risk tolerances, and is gradually becoming recognized as important in the space weather community \citep{Camporeale2019}. Additionally, human forecasters have greater trust in automated tools such as ML models when a forecast is accompanied by a reliable probability. There has been some research on probabilistic solar flare forecasting including \citet{Leka2007,wheatland2004bayesian}, a comparison of models on probabilistic metrics in \citet{Barnes2016comparisonI}, and the more recent work of \citet{Guerra2020,Nishizuka2020,Sun2022}. However, comparison to the literature has often relied on converting probabilistic forecasts to binary ones by applying a threshold value and many other works have focused on optimizing binary skill scores such as the True Skill Statistic (TSS).

The remainder of this paper is organized as follows. Section \ref{sec:data} discusses the data pipeline, Section \ref{sec:models} introduces the ML models, and Section \ref{sec:evaluation} outlines how we will evaluate our performance. In Section \ref{sec:featureimportance}, we use a simple logistic regression model to evaluate different scalar features. In Section \ref{sec:modelselection}, we conduct experiments to refine a CNN model, and compare three different ML models. Finally, in Section \ref{sec:histvshmi} we compare the performance of our final ML model when trained on the historical era to a model trained on the HMI era alone. Conclusions and suggestions for future work are given in Section \ref{sec:conclusion}.
	
\section{Data} \label{sec:data}

\subsection{Preprocessing and labeling}

\label{sec:preprocessing}
We consider magnetogram data from 5 different instruments spanning 1975-present: Mount Wilson Observatory (MWO){(available from \citet{mwo})}, KPVT/SPMG {(available from \citet{kpvtspmg}}, KPVT/512 {(available from \citet{kpvt512})}, SOHO/MDI and SDO/HMI\footnote{{MWO and KPVT data can be downloaded from the solar dynamo dataverse (\url{https://dataverse.harvard.edu/dataverse/solardynamo}), maintained by Andrés Muñoz-Jaramillo.}}. These data {exist} at varying cadence and resolution. In this work we take an approach of performing as little pre-processing as possible to synthesize the datasets.
We take daily samples and apply the following preprocessing procedure to all images (see example in Appendix \ref{appendix:preprocessing}):
\begin{enumerate}
    \item Apply a Gaussian kernel to smooth data to approximate 512x512 resolution ($\sigma = 4, 1, 1.74, 2$ for HMI, MDI, SPMG, and 512 accordingly)
    \item Rotate to align with world-coordinate axis.
    \item Reproject \citep{reproject} to a virtual instrument at 1au with 1024x1024 resolution and 2.2" plate scale.
    \item Zero out limbs past 95\% of solar radius.
\end{enumerate}

For MDI, we divide magnetograms by 1.3 before applying the procedure above, but do not apply any calibration factor to KPVT/512, KPVT/SPMG, and SDO/HMI.  This form of intercalibration is based on the results of \citet{Munoz-Jaramillo-etal2021}. For MWO, we multiply by a calibration factor of 2 and do not perform any smoothing. This rough calibration factor was chosen by comparing distributions of total unsigned flux during overlapping instrument periods, as seen in Appendix \ref{appendix:totusflux}. More rigorous comparisons have inferred conversion factors for mapping MWO to MDI of 0.5-6 \citep{Riley2014,Tran2005}.

\begin{figure}[htbp]
    \centering
    \includegraphics[scale=0.8]{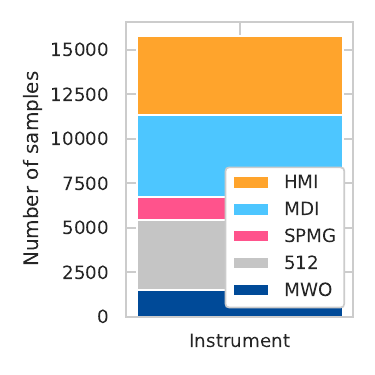}
    \includegraphics[scale=0.8]{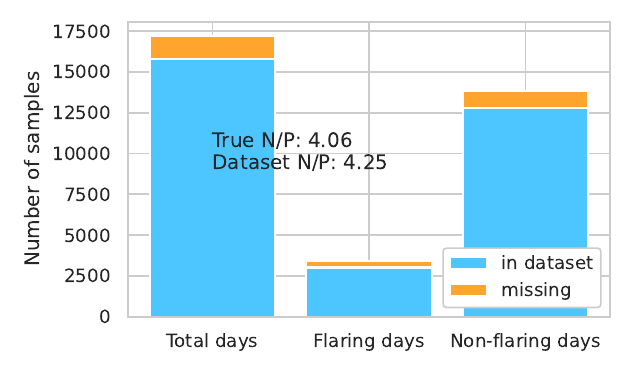}
    \caption{Left, breakdown of daily samples by instrument. Right, breakdown of flaring vs. non-flaring sample days contained in the dataset.}
    \label{fig:databreakdown}
\end{figure}

Taking a single daily sample according to instrument preference of HMI$>$MDI$>$SPMG$>$512$>$MWO results in a final dataset of 15,800 samples. {For HMI and MDI, which are available at a high temporal cadence, we take the closest available sample to 00:00 UT on each day.} Fig. \ref{fig:databreakdown} shows the breakdown of this dataset by instrument.

We use the Heliophysics Event Knowledgebase (HEK) flare catalog \citep{hurlburt2012heliophysics} queried through the SunPy Fido module to label our dataset \citep{sunpy_community2020}. Using a label of P $=$ M1 or larger flare within the next 24 hours and N $=$ no M1 or larger flares within the next 24 hours, we can see how many samples are flaring vs. non-flaring. We can also see how many days are captured within our dataset and how many samples are missing. Fig. \ref{fig:databreakdown} shows the breakdown of missing and non-missing daily samples in our dataset. Note that unlike many other works, we consider full disk magnetograms so we do not remove limb flares. {We zero out data past 95\% of the solar radius in order to handle issues with magnetogram measurements close to the limb. It would be of interest to analyze model predictive ability for limb flares compared to flares on the disk. However, given that our results in Section \ref{sec:histvshmi} are very similar to prior studies that have only used on-disk data, we do not consider this to be a significant source of variability in performance and we leave this analysis to future work.}

\subsection{Scalar Features} \label{sec:scalarfeatures}

We consider extracting scalar features from the single frame magnetograms to use as additional inputs to a ML flare forecasting model: total unsigned flux, total signed flux, minimum value, and maximum value. To compute total signed and unsigned flux, we use a minimum magnetic field of 30 Gauss to threshold the preprocessed magnetogram data and a minimum area of 64 pixels to reduce noise and only include flux from larger active regions.

Besides magnetogram-based features, it is known that including information about flare history can improve forecasts \citep{falconer2012prior, Jonas2018, wheatland2004bayesian}. We experiment with adding the following features to capture climatology: flaring rate over the past week, month and year, and largest flare magnitude in the last 24, 48 and 72 hours.
 
\subsection{Data splitting}
\label{sec:splitting}
Taking daily magnetogram samples from 1975-2022, we have a labeled dataset with 15,800 preprocessed samples. Each sample has an associated set of scalar features described in Section \ref{sec:scalarfeatures}, and the associated maximum flare intensity within a 24 hour forecast window. 

Dataset imbalance is a well-known problem in flare forecasting, and other machine-learning applications. Two common approaches for dealing with imbalanced data are using weighted loss functions and artificial balancing of the dataset, either through undersampling of the majority class or oversampling of the minority class. A weighted focal loss was used in \citet{deshmukh2022decreasing}, while several balancing approaches were explored in \citet{Ahmadzadeh_2021} in order to preserve climatology. Regardless of which approach is taken, there is a challenge in applying models to predictions on data which has a different distribution than the training data. This is especially relevant in solar flare forecasting where there is considerable variability in flaring both over the solar cycle and between different solar cycles. We note that the imbalance over the full historical dataset, as seen in Fig. \ref{fig:databreakdown}), is smaller than over the HMI era. In our experiments, we found that neither a weighted loss function nor an approach of undersampling the majority class across ensemble members as in \citet{ChatterjeeSEP} improved performance compared to doing nothing to address data imbalance. {Note that this is contrary to the findings of \citet{Ahmadzadeh_2021} where both balancing and weighted loss functions were found to improve performance in a binary classification setting for flare forecasting. It could be that this advantage is simply not seen in the probabilistic setting. However, it is more likely that we do not see an improvement simply because the dataset imbalance that we are dealing with (approximately 4:1 negatives to positives) is far less than the imbalance considered in \citet{Ahmadzadeh_2021} (ranging from 20:1 to 95:1).}

We split the dataset temporally in order to avoid temporal leakage and approximate an operational setting. For every year, we hold out samples in November and December as a first test set (Test A). We additionally withhold data from Jan 1, 2016 to Dec 31, 2017 in order to compare with the benchmarking done in \citet{Barnes2016comparisonI, Leka2019a, Leka2019b} (Test B). These test sets will remain unseen throughout all experimentation and model tuning. They will only be used for final evaluation.
In addition to the test sets, we remove samples between September 15 and October 31 of each year as an additional hold-out set. This will be used to make high-level decisions on the model and training process. The remainder of data is used for training and validation. Table \ref{tab:datasetclimatology} describes the flare climatology of these temporally split datasets. We can see that other than Test B, which is taken over a particularly quiet period as can be seen in Fig. \ref{fig:flareshistorical}, the climatology is similar across datasets. Although we wish to use Test B to compare to the results in \citet{Barnes2016comparisonI, Leka2019a, Leka2019b}, given that this 2016-17 time period is so quiet relative to the rest of the past 40 years, it is not a very representative test set. 

\begin{table}[hbt]
    \centering
    \begin{tabular}{lcccccc}
        Dataset & Forecast Window & \# X flares & \# M flares & \# C flares & \# No flares & N/P Ratio \\ \hline
        Test Set A (2016 - 2017) & \multirow{5}{*}{24 hours} & 3 & 24 & 184 & 518 & 27.2 \\
        Test Set B (Nov - Dec) &  & 59 & 535 & 1486 & 1140 & 3.76 \\
        Hold-out Set (Sep 15 - Oct) &  & 43 & 356 & 1089 & 905 & 4.48 \\
        Train \& validation Set (Jan - Sep 15) &  & 259 & 2003 & 5790 & 4783 & 4.15 \\
        Total & & 364 & 2916 & 8532 & 7156 & 4.25 \\
    \end{tabular}%
\caption{Flare climatology for each dataset given a 24 hour forecast window. The number of events (X flares, M flares, etc.) is defined as the number of samples which contain an event within the forecast window. We only count the largest event, i.e., if a sample contains both M and C flares within the forecast window, it will be counted as an M flare sample. The ratio of negatives to positives, N/P, is taken as the sum of C and no flare events divided by the sum of M and X flare events.}
\label{tab:datasetclimatology}
\end{table}

\begin{figure}[htbp]
    \centering
    \includegraphics[width=0.7\textwidth]{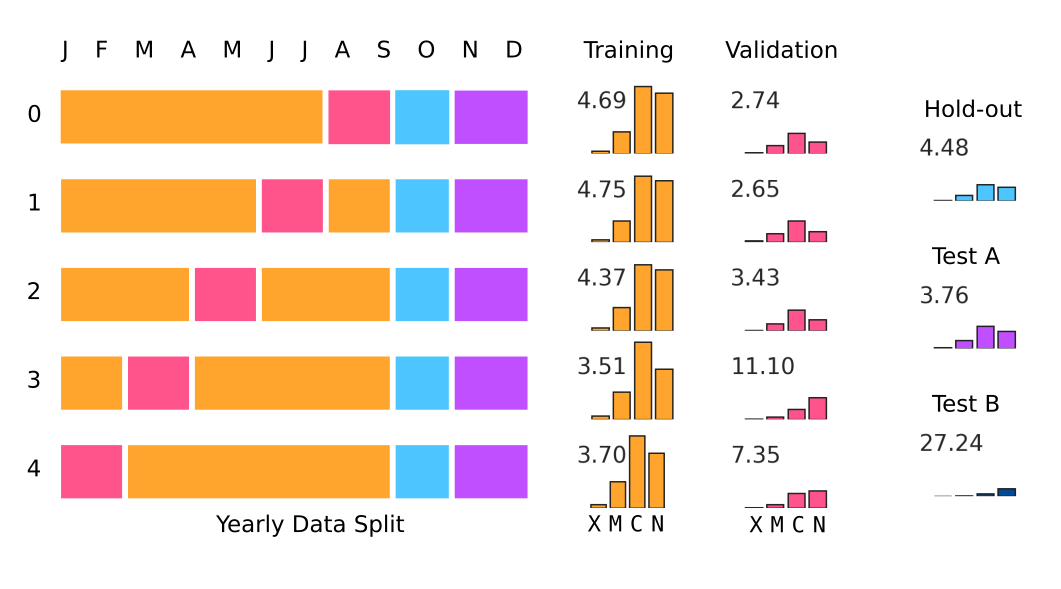}
    \caption{Yearly training, validation, hold-out and test splits for the 5 ensemble members (left), along with corresponding 24 hour flare climatology distributions and N/P ratios (right). Note that Test B is the 2016-2017 data, not depicted on the yearly split.}
    \label{fig:datasplit}
\end{figure}

We create multiple training and validation sets by splitting the data from January 1 to September 15 into 5 partitions and using each partition as a validation set with the remainder for training, as shown in Fig. \ref{fig:datasplit}. By taking the ensemble of these trained models, we can get a better evaluation of performance on the hold-out and generate probabilistic forecasts with uncertainty. Table \ref{tab:trainvalclimatology} gives the flare climatology for each of these splits.

\begin{table}[htb]
    \centering
    \begin{tabular}{llcccccc}
        Split & Dataset & Forecast Window & \# X flares & \# M flares & \# C flares & \# No flares & N/P Ratio \\ \hline
        0 & Training/Validation & \multirow{5}{*}{24 hours} & 184/75 & 1447/556 & 4451/1339 & 4014/769 & 4.69/2.74 \\
        1 & Training/Validation &  & 168/91 & 1433/570 & 4387/1403 & 4075/708 & 4.75/2.65 \\
        2 & Training/Validation &  & 211/48 & 1539/464 & 4401/1389 & 4053/730 & 4.37/3.43 \\
        3 & Training/Validation &  & 234/25 & 1836/167 & 5116/674 & 3346/1437 & 3.51/11.1 \\
        4 & Training/Validation &  & 239/20 & 1757/246 & 4805/985 & 3644/1109 & 3.70/7.35 \\
    \end{tabular}%
    \caption{Flare climatology for the different training and validation splits.}
    \label{tab:trainvalclimatology}
\end{table}

\begin{figure}[hbt]
    \centering
    \includegraphics[width=0.85\textwidth]{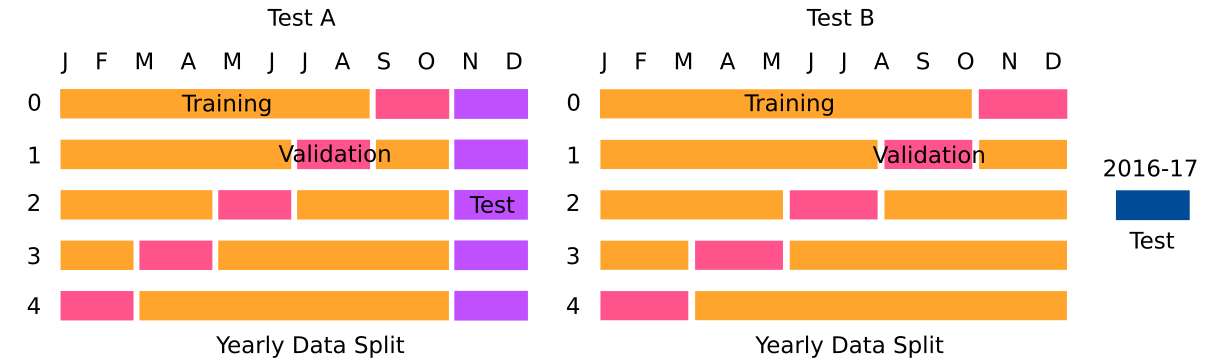}
    \caption{Yearly training, validation, and test splits for the 5 ensemble members for both evaluation test cases.}
    \label{fig:datasplittest}
\end{figure}

After making all model decisions, we will retrain models including the hold-out set in the training and validation. For evaluating on Test A, we will also include the data from 2016-2017 in training. For evaluating on Test B, we will include the November-December data for training. In both cases we remove 5 days on either side of the test data from training/validation in order to prevent temporal leakage. To measure the effect of including historical data, we will train and evaluate models using all instruments, and compare to models trained and evaluated on only HMI data. In either case, the same temporal splitting strategy will be used. Fig. \ref{fig:datasplittest} illustrates the temporal splitting for evaluation on Test A and Test B.

\section{Models} \label{sec:models}

For all models, we will use an ensemble approach where predictions are made using the median output of the ensemble members. We train 5 ensemble members on different subsets of data, as described in Section \ref{sec:data}. In general, the performance of the ensemble-median is expected to be better than the performance of the individual members. 

\subsection{Logistic Regression} \label{sec:lr}

The logistic regression (LR) model uses a sigmoid function to model the probability of a binary output as a function of weighted input features. The sigmoid function can be written as 
\begin{equation}
    h_\Theta = \frac{1}{1+e^{-\Theta_1^T x -\Theta_0}}
\end{equation}
where $x$ is the input feature vector and $\Theta$ are the weights to be learned. The LR model can be interpreted as modeling the output variable as a Bernoulli distribution conditioned on the input features, where the weights are estimated from training data using Maximum Likelihood Estimation (MLE). This is a convex optimization problem, thus has a unique solution. We use the LR implementation in \texttt{sklearn} \citep{scikit-learn} and the LBFGS algorithm for optimization.

Despite their simplicity, LR models have been used for flare forecasting \citep{Aktukmak2022,Pandey2022,Song2009,Yuan2010} along with other generalized linear models such as Lasso \citep{Jonas2018} and Support Vector Machines \citep{Bobra2015,Nishizuka2017,Yuan2010} with comparable skill to more complex models.

\subsection{Convolutional Neural Network} \label{sec:cnn}

We adapt the CNN from \citet{ChatterjeeSEP} as our machine learning model. As input, we consider a single full-disk magnetogram. We resize the input to 128x128 and clip maximum pixel values to $\pm 1000$ {G}. {Note that we did not find any improvement in performance by increasing our input resolution to 256x256, indicating that the CNN is not learning more relevant information from higher spatial resolution. Although 128x128 resolution may seem coarse for full disk images, this is an input size of 16,384 features, which is over two orders of magnitude larger than the number of scalar features computed from higher resolution magnetograms, i.e., the SHARPs parameters \citep{bobra:2014}, often used as input to flare forecasting models}. The output is a single value between 0 and 1, representing the probability of flaring within a given forecast window. 

For training we use the binary cross entropy loss function and Adams optimizer with cosine annealing to vary the learning rate \citep{Loshchilov2017}. We select reasonable choices for hyperparameters based on initial experimentation and common practice: batch size 256, learning rate $10^{-4}$, weight decay $10^{-3}$. In all cases, we train for a maximum of 100 epochs and select the best model as the one with the lowest validation loss.

\begin{figure}[hbtp]
    \centering
    \includegraphics[width=\textwidth]{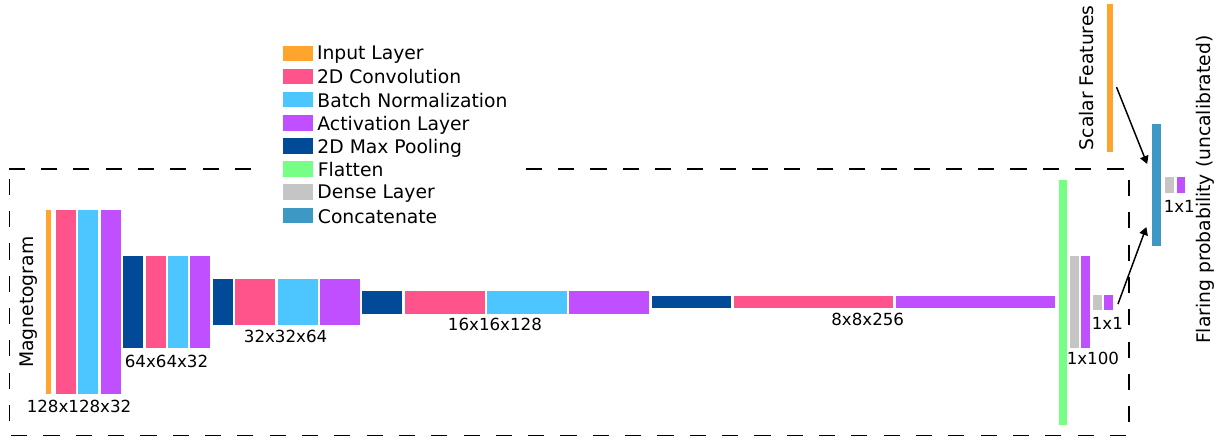}
    \caption{CNN+LR architecture, where the dashed black box indicates the CNN alone. A 128x128 magnetogram is fed through a series of 2-D convolutions (all kernels of size 3x3), batch normalization, non-linear activation, and max pooling. The flattened output is then fed through two dense layers to produce a probability of flaring. When scalar features are included, they are concatenated with this 1x1 CNN output and fed through another dense layer to obtain a final flaring probability.}
    \label{fig:cnnarchitecture}
\end{figure}

In Section \ref{sec:modelselection} we consider augmenting the CNN with an additional fully connected layer to incorporate scalar features. We initialize this layer with weights obtained from the LR model, and then retrain the full architecture. We refer to this architecture as the CNN+LR model. Fig. \ref{fig:cnnarchitecture} depicts this model architecture, including the CNN only portion. 

\section{Performance evaluation} \label{sec:evaluation}

\subsection{Metrics} \label{sec:metrics}
To evaluate performance, we focus on probabilistic metrics rather than metrics derived from setting a binary threshold and computing a contingency table. We consider mean squared error (MSE) and the brier skill score (BSS) between the predicted probabilities and the true labels { defined as:}

\begin{align}
    &\text{MSE}(y_p,y) = \langle(y_p - y)^2\rangle, \\
    &\text{BSS} = \frac{\text{MSE}(y_p,y)-\text{MSE}(\langle y \rangle,y)}{0-\text{MSE}(\langle y \rangle,y)},
\end{align}

{where $y_p$ is the predicted probability, $y$ is the true outcome (0 for no-flare and 1 for flare), and $<>$ is the averaging operator. }
MSE, otherwise known as the Brier Score, ranges from $[0,\inf)$ with 0 indicating a perfect forecast. BSS is a mean-square-error skill score measured against the reference of the climatological forecast of the evaluation period, where 1 is the maximum perfect score, 0 is no skill added, and negative values indicate worse performance than climatology. Note that MSE is strongly affected by dataset imbalance, hence, the MSE value can only be compared on the same test dataset. 

Forecasts can also be summarized with precision-recall curves and receiver operating characteristic (ROC) curves. The average precision score (APS) is a measure of the area under the precision-recall curve with a maximum perfect score of 1. The Gini coefficient, or ROC skill score is a normalized version of the area under the ROC-curve where no skill results in a score of 0 and perfect forecasts give a score of 1.

In order to assess whether a model is outputting reasonable probabilities, we consider the concept of reliability. A model is reliable if forecast probabilities are equal to the actual likelihood of an event, i.e., if the model outputs 60\%, there will be a flare 60\% of the time. In practice reliability can be measured by partitioning classifier probabilities into bins and measuring the difference from ground truth likelihoods for each bin. This can be visualized in a reliability diagram plotting true probability vs. predicted probability. We calculate two metrics in order to summarize reliability: expected calibration error (ECE) and maximum calibration error (MCE). ECE is the average deviation between predicted and true probability, while MCE is the maximum deviation{ defined as:}

\begin{align}
    \text{ECE} &= \frac{1}{\sum_{m=1}^M n_m} \sum_{m=1}^M n_m |P^{true}_m-P^{pred}_m| \\
    \text{MCE} &= \max_m |P^{true}_m-P^{pred}_m| 
\end{align}
{where given a reliability diagram obtained by binning predictions into $M$ bins, $P^{true}_m$ is the true frequency of positives, $P^{pred}_m$ is the average predicted probability, and $n_m$ is the total number of samples in bin $m$.}

\subsection{Probability Calibration} \label{sec:calib}

Model outputs can be calibrated through a post processing step in order to improve reliability. We consider Bayesian Binning Quantiles (BBQ) for probability calibration \citep{Naeini2015}. This non-parametric method uses a weighted combination of equal-frequency histogram binning models with different numbers of bins. The weights are proportional to the corresponding Bayesian Score given a prior Beta distribution over binning model parameters. As in \citet{ChatterjeeSEP}, we calibrate on the combined training and validation set. Each member of the ensemble is calibrated, and the final calibrated ensemble-median taken as the median output of the individually calibrated models.

\section{Feature Importance} \label{sec:featureimportance}

\begin{figure}[htbp]
    \centering
    \includegraphics[width=0.48\textwidth]{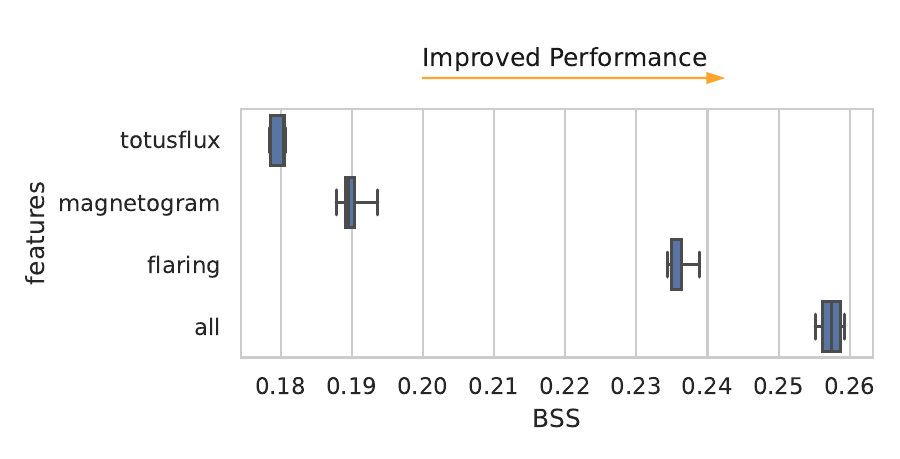}
    \includegraphics[width=0.48\textwidth]{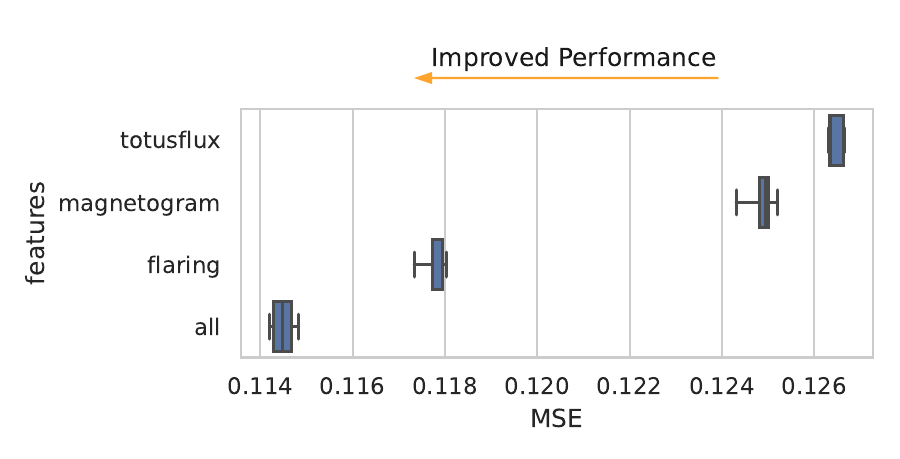}
    \includegraphics[width=0.48\textwidth]{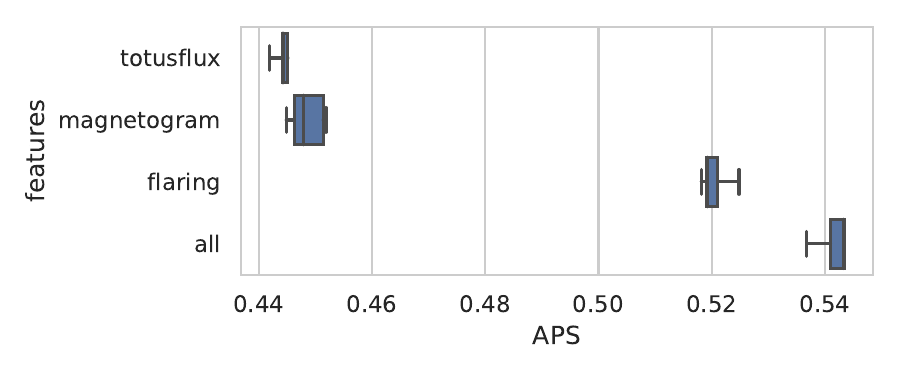}
    \includegraphics[width=0.48\textwidth]{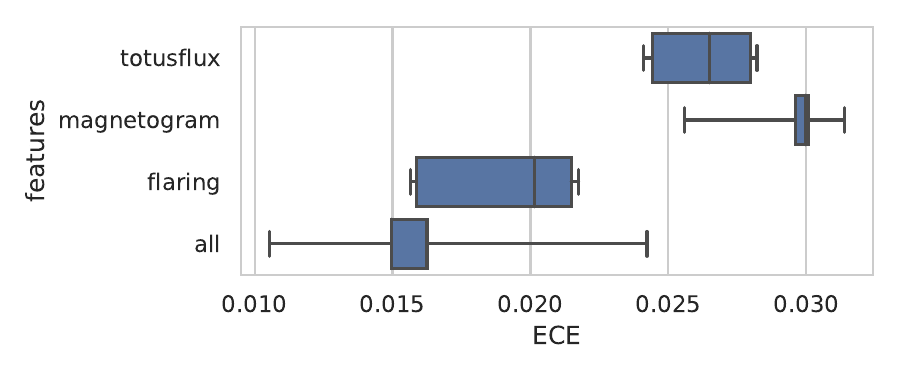}
    \includegraphics[width=0.48\textwidth]{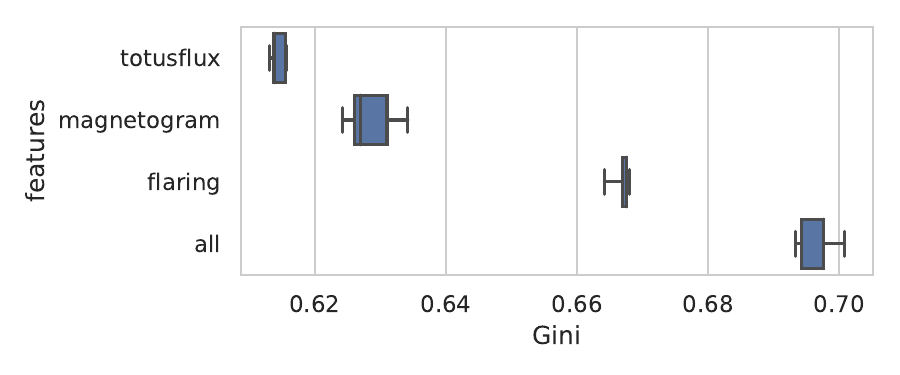}
    \includegraphics[width=0.48\textwidth]{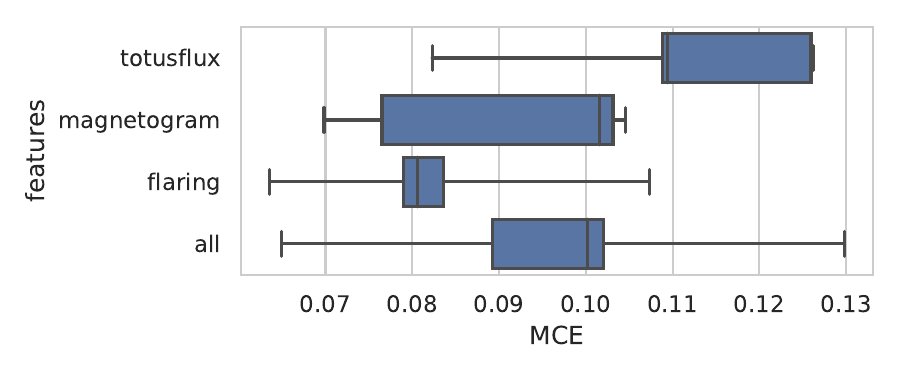}
    \caption{Comparison of metrics evaluated on the hold-out set using the probability calibrated LR models on varying input feature sets. Left, BSS, APS and Gini index are metrics to be maximized with optimal values of 1. Right, MSE, ECE and MCE are metrics to be minimized with optimal values of 0.}
    \label{fig:LR_featurecomp}
\end{figure}

We consider training the logistic regression ensemble on sets of scalar features in order to get a sense of how predictive different features are for flaring. We consider models trained on the total unsigned flux alone, all magnetogram derived features (total unsigned flux, total signed flux, data minimum and data maximum), all flaring history related features (flaring rate over the past week, month, year and maximum flare in the past 24, 48 and 72 hours), and the combined set of all features. Fig. \ref{fig:LR_featurecomp} compares the performance of the LR ensemble members on the hold-out set given the different input feature sets, while Table \ref{tab:LR_featurecomp} summarizes the performance of the ensemble. In general, the ensemble members perform very similarly. We can see that including all features results in the best performance across all metrics except MCE where performance is similar across feature sets. Interestingly, the flaring history features perform better than the magnetogram based features. This indicates that single frame magnetogram inputs have limited predictive power. 

\begin{table}[htbp]
    \centering	
    \begin{tabular}{lcccccc}
        Feature Set & MSE$_\downarrow$ & BSS$_\uparrow$ & APS$_\uparrow$ & Gini$_\uparrow$ & ECE$_\downarrow$ & MCE$_\downarrow$ \\ \hline
        Total US Flux & 0.13 (0.00) & 0.18 (0.00) & 0.45 (0.09) & 0.61 (0.00) & 0.03 (0.00) & 0.11 (0.02)\\
        Magnetogram & 0.12 (0.00) & 0.19 (0.00) & 0.45 (0.00) & 0.63 (0.00) & 0.03 (0.00) & 0.10 (0.02)\\
        Flare History & 0.12 (0.00) & 0.24 (0.00) & 0.52 (0.00) & 0.67 (0.00) & 0.02 (0.00) & 0.07 (0.02)\\
        All & 0.11 (0.00) & 0.26 (0.00) & 0.55 (0.01) & 0.70 (0.00) & 0.02 (0.00) & 0.11 (0.02)\\
    \end{tabular}
    \caption{Performance of the calibrated ensemble-median for a 24 hour flare forecast on the hold-out set using different feature sets and the LR model. The standard deviation of the ensemble members is given in parentheses. {Subscript arrows indicate the direction of improvement for the given metric.}}
    \label{tab:LR_featurecomp}
\end{table}

\begin{figure}[hbtp]
    \centering
    \includegraphics[width=0.5\textwidth]{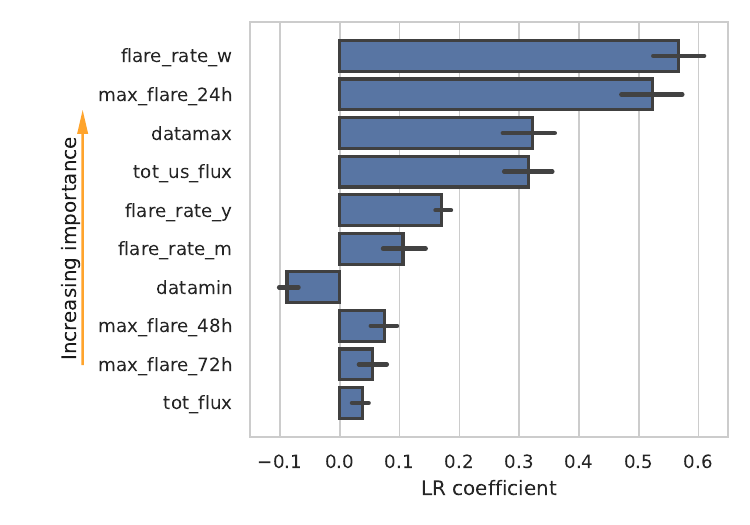}
    \caption{Coefficients of the logistic regression model trained on all features. Bias is the constant term in the model. The plot shows the mean value of the coefficient for each feature across all models, with the error bars for 95\% confidence intervals.}
    \label{fig:LR_coeffs}
\end{figure}

Looking at the final coefficients of the trained LR model is one way to quantify how important each feature is for this particular model since all the input features are standardized. Note that this is not necessarily a measure of how predictive each feature is in general. Fig. \ref{fig:LR_coeffs} shows the coefficients of the trained LR ensemble using all input features. Even though the ensemble is made up of 5 models trained on different subsets of data, the model coefficients are fairly consistent, as can be seen from the confidence intervals.

\section{Model Selection} \label{sec:modelselection}

\subsection{Pre-training}
\label{sec:pretraining}

Ultimately, we seek to train our model on a binary output of flare or no-flare within a specified forecast window, with the threshold for flaring being M1 and larger flares. Given the correlation of total unsigned flux with flaring, we consider using this quantity as a proxy label to pre-train our models. In particular, we consider 3 different methods:
\begin{enumerate}
    \item No pre-training: Training from scratch to forecast flare/no-flare
    \item Pre-training A: Using a flux threshold of $9e6$ $G Mm^2$ 
    \item Pre-training B: Using a flux-dependent flare threshold $$\log(\text{max flare intensity}) > -3 - \frac{2}{1.5e7 G Mm^2} \text{total unsigned flux}$$
\end{enumerate}

When pre-training, we start by using the binary label defined for scheme A or B, train the model until the validation loss stops decreasing, and then use those weights as the initial weights for training on the flare/no-flare label. 

We train 5 different models, one for each training/validation split, and evaluate on the hold-out set in order to determine which method is best. The results for a 24 hour forecast are given in Table \ref{tab:24hpretraining}. Overall, we see that pre-training improves performance, with the high-flux threshold giving the most reliable improvement. In this experiment we find that probability calibration decreases performance, so we report the results from the uncalibrated models.

\begin{table}[htbp]
    \centering	
    \begin{tabular}{lcccccc}
        Experiment & MSE$_\downarrow$ & BSS$_\uparrow$ & APS$_\uparrow$ & Gini$_\uparrow$ & ECE$_\downarrow$ & MCE$_\downarrow$ \\ \hline
        No pre-training & 0.12 (0.00) & 0.20 (0.01) & 0.47 (0.01) & 0.66 (0.01) & 0.03 (0.00) & 0.25 (0.07)\\
        Pretraining A & 0.12 (0.00) & 0.21 (0.01) & 0.49 (0.02) & 0.66 (0.01) & 0.03 (0.01) & 0.13 (0.09)\\
        Pretraining B & 0.12 (0.00) & 0.19 (0.01) & 0.45 (0.01) & 0.64 (0.02) & 0.03 (0.01) & 0.38 (0.13)\\
    \end{tabular}
    \caption{Uncalibrated ensemble-median results on the hold-out set using different pre-training strategies. {Subscript arrows indicate the direction of improvement for the given metric.}}
    \label{tab:24hpretraining}
\end{table}

\subsection{Data augmentation}
\label{sec:augmentation}

Data augmentation is a widely used technique in machine learning. It can be used to address data imbalance, by augmenting the minority class, or just to reduce overfitting by introducing randomness in extending the dataset. Augmentation has been used in previous flare-forecasting studies, with mixed results \citep{deshmukh2022decreasing,Guastavino2022,Li2020}. Here we consider using data augmentation on all training data to increase generalizability, not to address dataset imbalance. We test three different approaches:

\begin{enumerate}
    \item No augmentation
    \item Conservative augmentation including random polarity reversals and random vertical flips
    \item Full augmentation including random polarity reversals, random vertical flips, random horizontal flips, and random rotations up to 20$^{\circ}$
\end{enumerate}

Based on the results of Section \ref{sec:pretraining}, we perform this experiment only on the 24 hour forecast using high-flux pre-training. The results are summarized in Table \ref{tab:augmentation}. When augmentation is used, probability calibration improves performance so here we report calibrated results.
We can see that augmentation is beneficial so we use the conservative augmentation approach for the remainder of our experiments.

\begin{table}[hbtp]
    \centering	
    \begin{tabular}{lcccccc}
        Experiment & MSE$_\downarrow$ & BSS$_\uparrow$ & APS$_\uparrow$ & Gini$_\uparrow$ & ECE$_\downarrow$ & MCE$_\downarrow$\\ \hline
        No augmentation & 0.12 (0.00) & 0.19 (0.03) & 0.48 (0.02) & 0.66 (0.04) & 0.04 (0.01) & 0.19 (0.10)\\
        Conservative augmentation & 0.12 (0.00) & 0.22 (0.01) & 0.50 (0.01) & 0.67 (0.01) & 0.02 (0.01) & 0.11 (0.04)\\
        Full augmentation & 0.12 (0.00) & 0.23 (0.01) & 0.50 (0.01) & 0.68 (0.01) & 0.03 (0.01) & 0.11 (0.04)\\
    \end{tabular}
    \caption{Calibrated ensemble-median results on the hold-out set using different amounts of data augmentation in training. {Subscript arrows indicate the direction of improvement for the given metric.}}
    \label{tab:augmentation}
\end{table}

\subsection{Model comparison}
\label{sec:modelcomparison}

\begin{figure}[hbtp]
    \centering
    \includegraphics[width=0.48\textwidth]{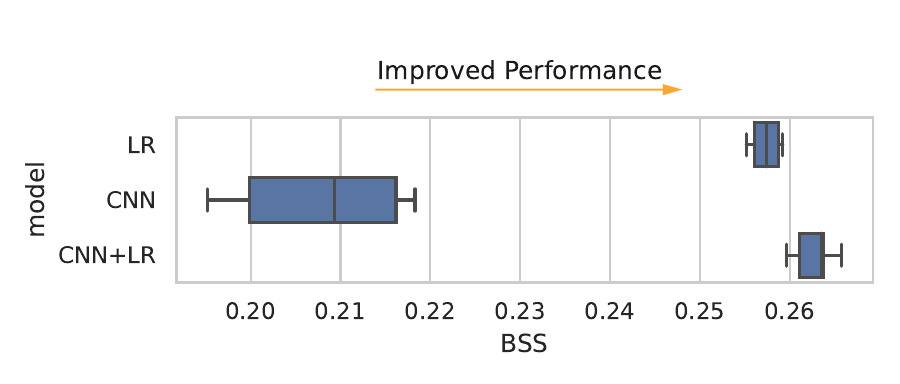}
    \includegraphics[width=0.48\textwidth]{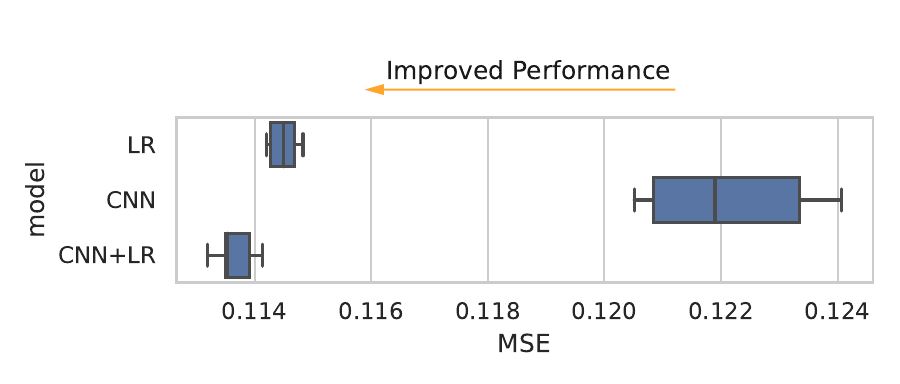}
    \includegraphics[width=0.48\textwidth]{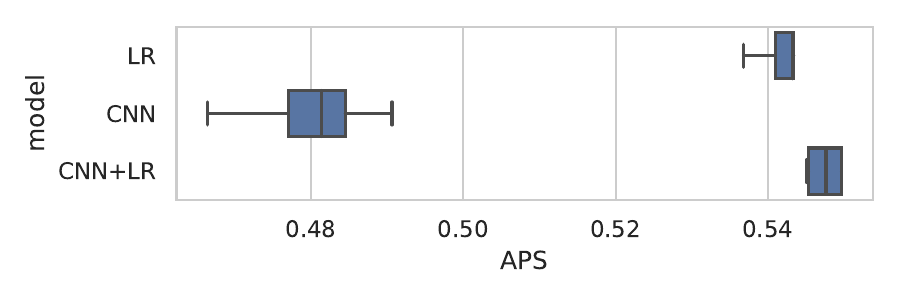}
    \includegraphics[width=0.48\textwidth]{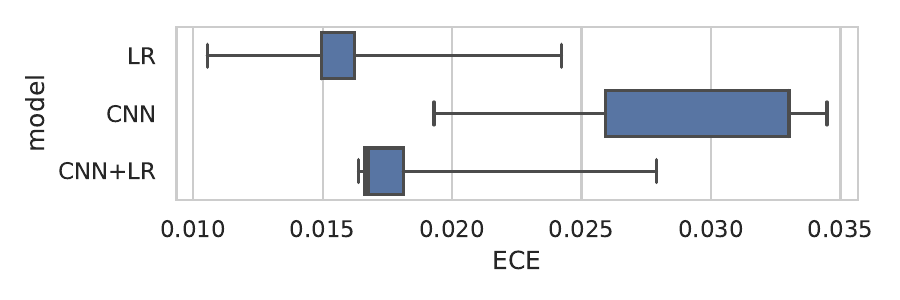}
    \includegraphics[width=0.48\textwidth]{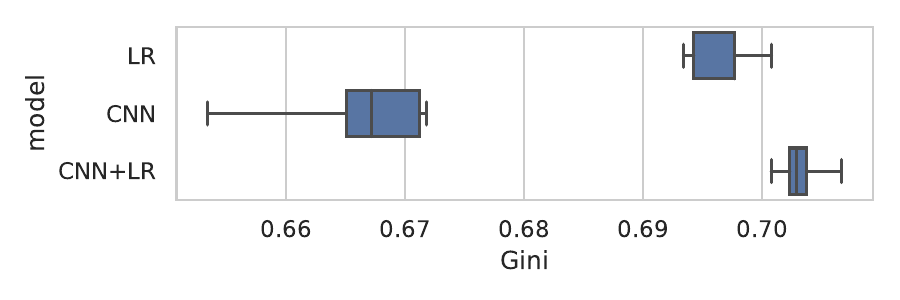}
    \includegraphics[width=0.48\textwidth]{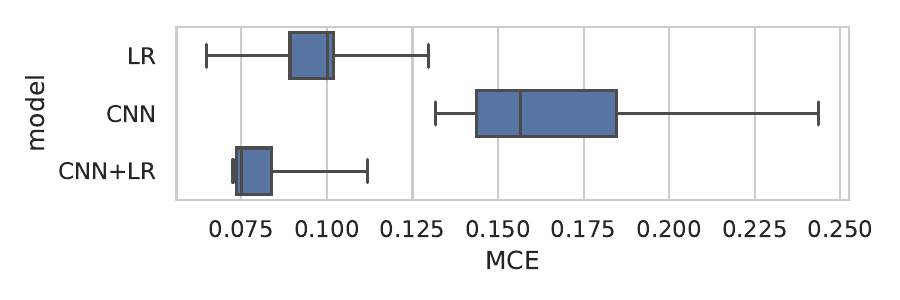}
    \caption{Comparison of metrics evaluated on the hold-out set using the different probability calibrated models. Left, BSS, APS and Gini index are metrics to be maximized with optimal values of 1. Right, MSE, ECE and MCE are metrics to be minimized with optimal values of 0.}
    \label{fig:cnnvslr}
\end{figure}

Finally, we compare the performance of the best LR, CNN and CNN+LR models on the hold-out set. The LR model uses all input features. The CNN and CNN+LR models are pre-trained using the high-flux label and conservative data augmentation is applied. The CNN+LR model uses all input features and the weights from the trained LR model to initialize the final layer. Table \ref{tab:cnnvslr} summarizes the results of the ensemble-median for each model, while Fig. \ref{fig:cnnvslr} shows the performance across ensemble members.

The combination of CNN+LR has the best performance, however, it is only marginally better than the LR alone. Although the CNN using single frame magnetograms as input does have some skill in flare forecasting, it performs significantly worse than both the LR alone and the CNN+LR. This implies that the scalar features used as input to a simple LR model account for much more of the predictive power of the CNN+LR model. Although we only consider one specific CNN model, it appears that single frame magnetograms do not contain significantly more relevant information for flare forecasting than can be summarized in a small number of scalar features. The CNN model performs slightly better than the LR model trained on only total unsigned flux, or magnetogram-based features, but worse than the LR model trained on flare history features or all features (see Table \ref{tab:LR_featurecomp}).

\begin{table}[htbp]
    \centering	
    \begin{tabular}{lccccccc}
        Experiment & MSE$_\downarrow$ & BSS$_\uparrow$ & APS$_\uparrow$ & Gini$_\uparrow$ & ECE$_\downarrow$ & MCE$_\downarrow$ & Max Output$_\uparrow$\\ \hline
        LR & 0.11 (0.00) & 0.26 (0.00) & 0.55 (0.00) & 0.70 (0.00) & 0.02 (0.00) & 0.11 (0.02) & 0.84 (0.01)\\
        CNN & 0.12 (0.00) & 0.22 (0.01) & 0.50 (0.01) & 0.67 (0.01) & 0.02 (0.01) & 0.11 (0.04) & 0.82 (0.05)\\
        CNN + LR & 0.11 (0.00) & 0.27 (0.00) & 0.56 (0.00) & 0.71 (0.00) & 0.02 (0.00) & 0.09 (0.02) & 0.84 (0.01)\\
    \end{tabular}
    \caption{Probability calibrated ensemble-median results for a 24 hour flare forecast on the hold-out set comparing models. {Subscript arrows indicate the direction of improvement for the given metric.}}
    \label{tab:cnnvslr}
\end{table}

\begin{figure}[hbtp]
    \centering
    \includegraphics[width=\textwidth]{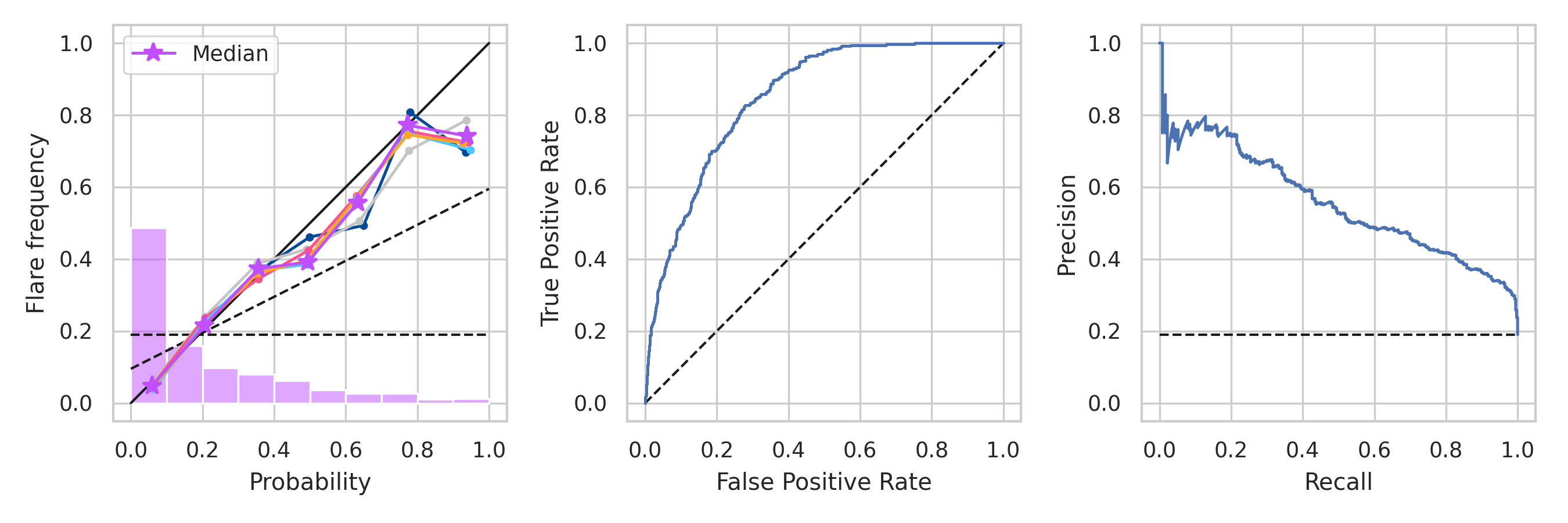}
    \includegraphics[width=\textwidth]{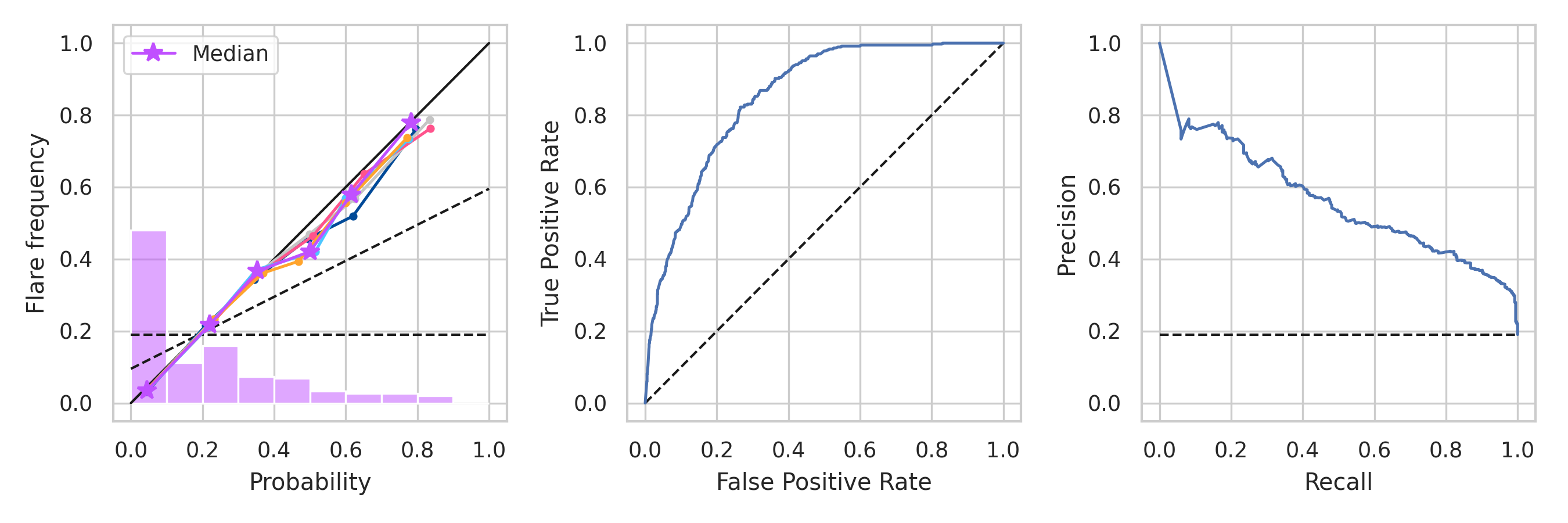}
    \caption{Reliability diagram, ROC-curve and precision-recall curve for the uncalibrated (top) and calibrated (bottom) CNN+LR ensemble using all input features and evaluated on the hold-out set. In the reliability diagram on the right, the dashed horizontal line indicates the climatological flare rate, while the dashed diagonal line indicates no-skill and the solid line indicates perfect reliability. In the ROC and precision-recall curves, the dashed lines indicate no-skill.}
    \label{fig:CNNLR_reliabilitydiag}
\end{figure}

The reliability diagram, ROC-curve and precision-recall curve for the CNN+LR model is shown in Fig. \ref{fig:CNNLR_reliabilitydiag} both before and after calibration. We can see that calibration improves reliability, while having no significant effect on the ROC-curve or precision-recall curve. Note that the calibrated probabilities reach a maximum around 0.85, indicating that the model is unable to give reliable forecasts with higher probability than this. This reduction in confidence is likely caused by the data imbalance and is a trade-off in optimizing for reliability.  

\section{Historical vs. HMI era} \label{sec:histvshmi}

To address the value of including historical data, we consider evaluation of trained models on two different test sets. The first is data from Nov-Dec of each year (Test A). The second is data from 2016-2017, which will evaluate performance on a relatively low flaring time period (Test B) and offer a comparison to \citet{Barnes2016comparisonI,Leka2019a,Leka2019b}. We might expect more training data to improve performance on the first test set, but not necessarily on the second, where the distribution is very different from the rest of the data. The first model is trained on the full historical dataset. The second is trained only on HMI data. The ML architecture for all cases is the CNN+LR model chosen based on the results of Section \ref{sec:modelselection}.

\subsection{Test A}\label{sec:testa}

\begin{figure}[hbtp]
    \centering
    \includegraphics[width=\textwidth]{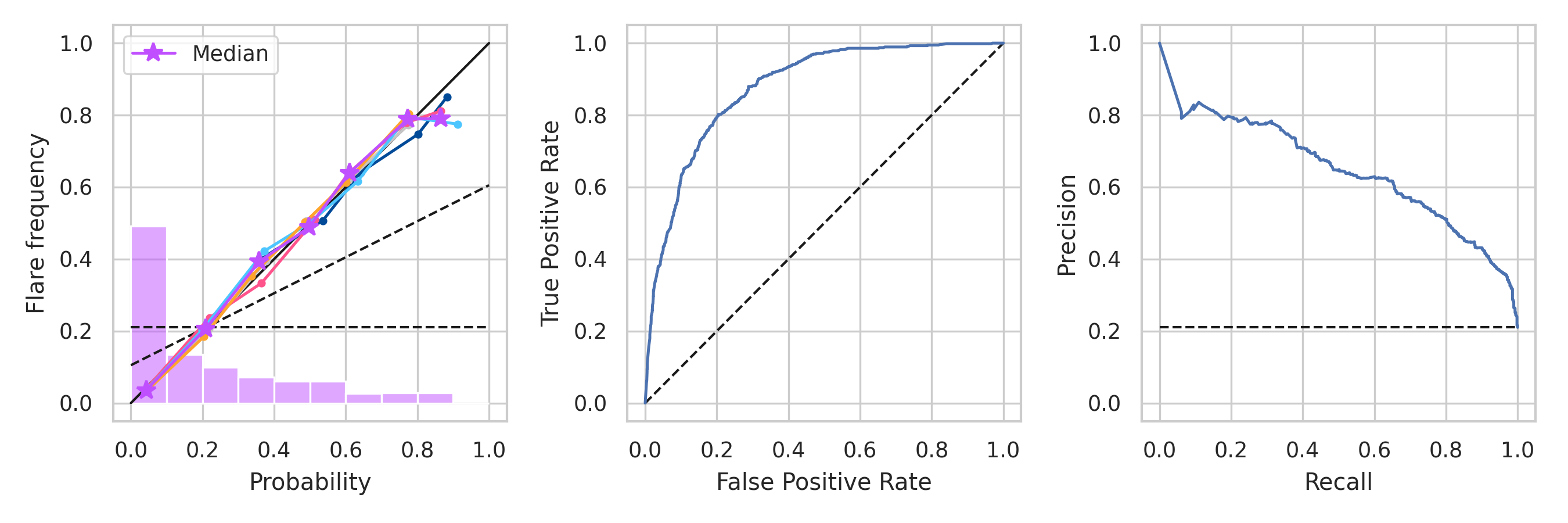}
    \includegraphics[width=\textwidth]{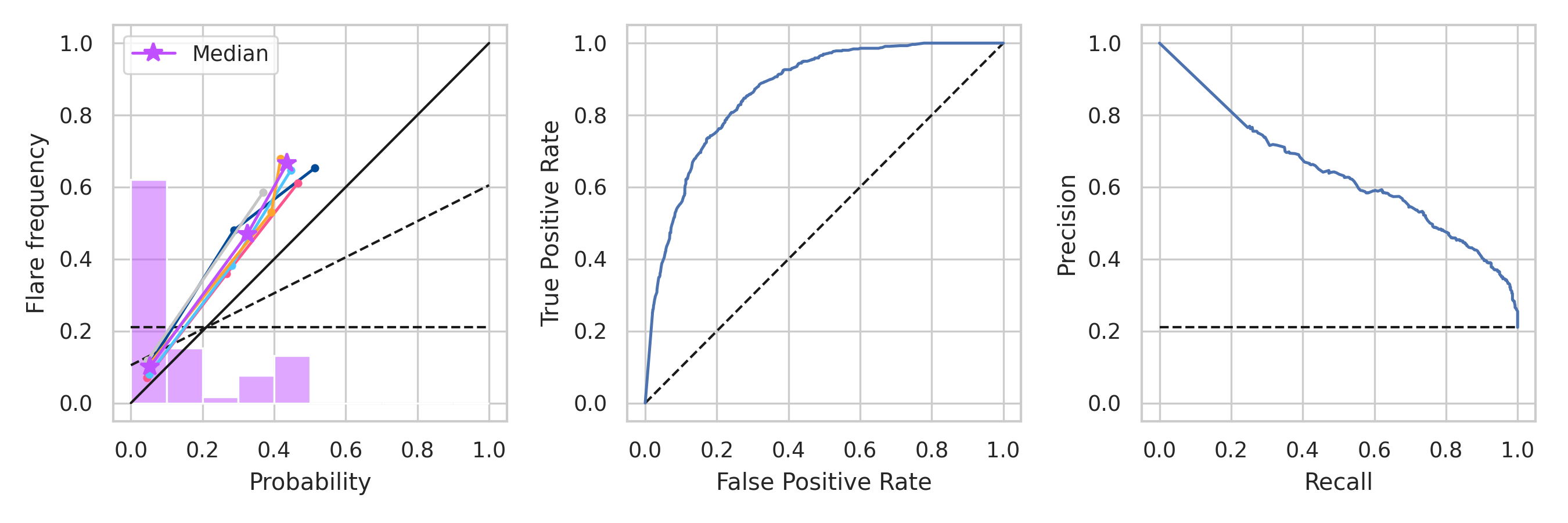}
    \caption{Performance of the calibrated CNN+LR ensemble on historical (top) and HMI (bottom) data, evaluated on Test A over the full timespan (Nov-Dec 1975-2022). Left, reliability diagram for the ensemble members and ensemble median. The dashed horizontal line indicates the climatological flare rate, the dashed diagonal line indicates no-skill and the solid line indicates perfect reliability. The histogram shows the distribution of output probabilities for the ensemble median. Center, the ROC curve and left, the precicison-recall curve for the calibrated ensemble median with no-skill indicated by a dashed line.}
    \label{fig:testa_performance_all}
\end{figure}

\begin{figure}[hbtp]
    \centering
    \includegraphics[width=\textwidth]{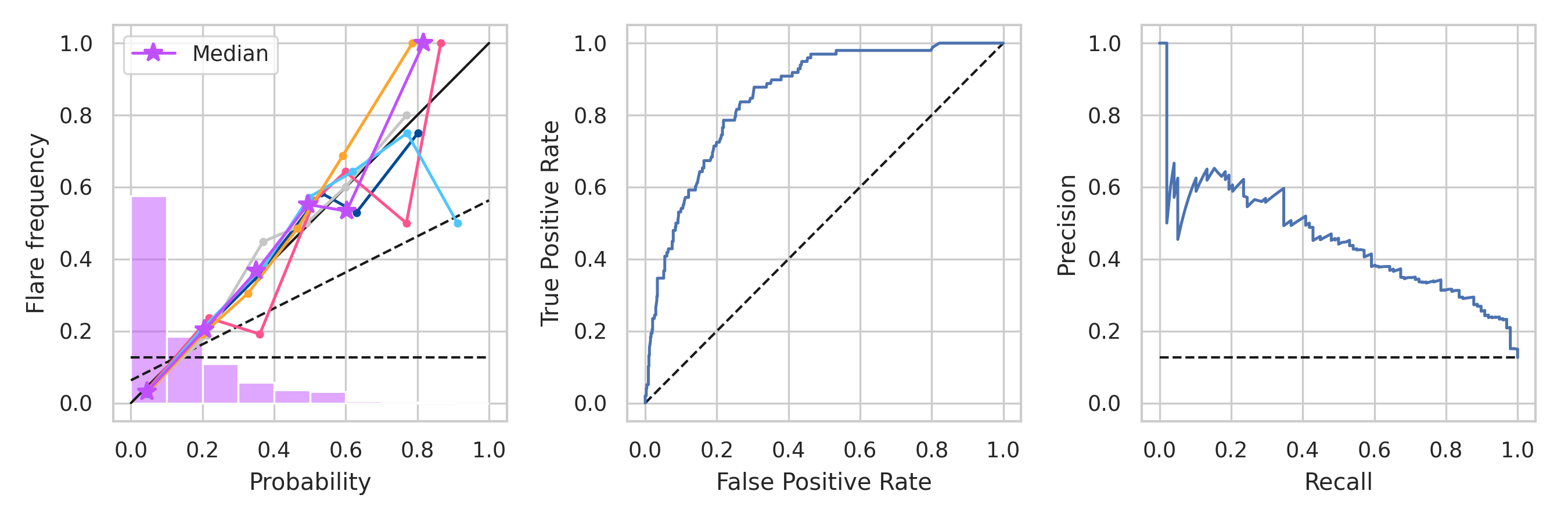}
    \includegraphics[width=\textwidth]{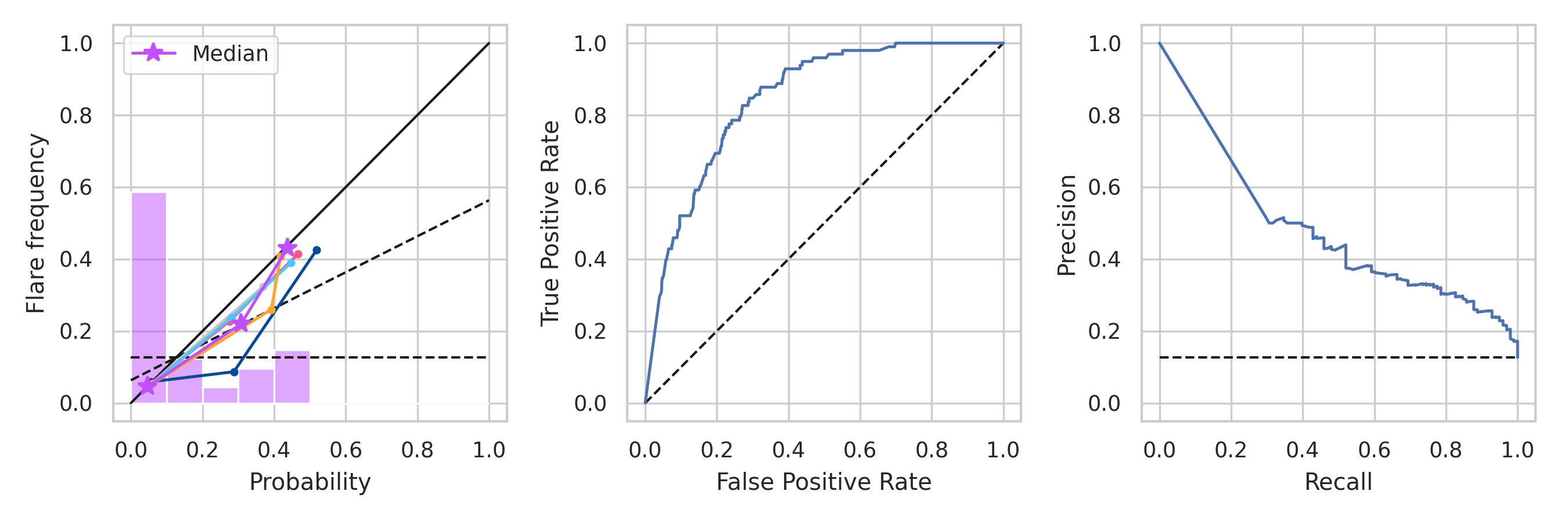}
    \caption{Performance of the calibrated CNN+LR ensemble on historical (top) and HMI (bottom) data, evaluated on Test A over the HMI era (Nov-Dec 2010-2022). Left, reliability diagram for the ensemble members and ensemble median. The dashed horizontal line indicates the climatological flare rate, the dashed diagonal line indicates no-skill and the solid line indicates perfect reliability. The histogram shows the distribution of output probabilities for the ensemble median. Center, the ROC curve and left, the precicison-recall curve for the calibrated ensemble median with no-skill indicated by a dashed line.}
    \label{fig:testa_performance}
\end{figure}

The model ensemble results for Test A are given in Table \ref{tab:testa}. We can see that the model trained on historical data generally outperforms the model trained on HMI data, especially when evaluated on the full dataset from 1975-2022, as seen in Fig. \ref{fig:testa_performance_all}. We additionally consider evaluation only on the modern HMI era from 2010-2022, as seen in Fig. \ref{fig:testa_performance}. Here the historical model performs better than the HMI-only model except on reliability metrics ECE and MCE. Looking at the reliability diagrams shown in Fig. \ref{fig:testa_performance} gives a fuller picture. Although the calibration errors may be lower in the HMI-only model, the quality of the calibrated forecast is extremely limited when trained on only HMI era data. The maximum output probability is below 0.5, which is not very useful for issuing confident flaring forecasts. 

\begin{table}[htbp]
    \centering	
    \begin{tabular}{llccccccc}
        Training & Testing & MSE$_\downarrow$ & BSS$_\uparrow$ & APS$_\uparrow$ & Gini$_\uparrow$ & ECE$_\downarrow$ & MCE$_\downarrow$ & Max Output$_\uparrow$\\ \hline
        Historical & Nov-Dec 1975-2022 & 0.11 (0.00) & 0.34 (0.01) & 0.64 (0.01) & 0.75 (0.01) & 0.01 (0.00) & 0.10 (0.03) & 0.87 (0.03)\\
        HMI & Nov-Dec 1975-2022 & 0.13 (0.01) & 0.23 (0.04) & 0.61 (0.02) & 0.73 (0.01) & 0.08 (0.02) & 0.33 (0.04) & 0.45 (0.08)\\ \hline
        Historical & Nov-Dec 2010-2022 & 0.09 (0.00) & 0.22 (0.01) & 0.46 (0.01) & 0.71 (0.01) & 0.02 (0.01) & 0.18 (0.07) & 0.83 (0.04)\\
        HMI & Nov-Dec 2010-2022 & 0.09 (0.00) & 0.18 (0.02) & 0.41 (0.03) & 0.69 (0.02) & 0.01 (0.01) & 0.12 (0.06) & 0.45 (0.08)\\
    \end{tabular}
    \caption{Probability calibrated ensemble-median results for a 24 hour flare forecast on Test A (Nov-Dec). The historical CNN+LR model is trained on data over all instruments from 1975-2022. The HMI CNN+LR model is trained only on SDO/HMI data available from 2010-2022. {Subscript arrows indicate the direction of improvement for the given metric.}}
    \label{tab:testa}
\end{table}

Looking at the results for 2010-2022 in Fig. \ref{fig:testa_performance} compared to the full duration 1975-2022 in Fig. \ref{fig:testa_performance_all}, we can see that the performance of the model trained on historical data does drop when evaluated on data with a significantly different flaring distribution. 

\subsection{Test B}\label{sec:testb}

\begin{figure}[htbp]
    \centering
    \includegraphics[width=\textwidth]{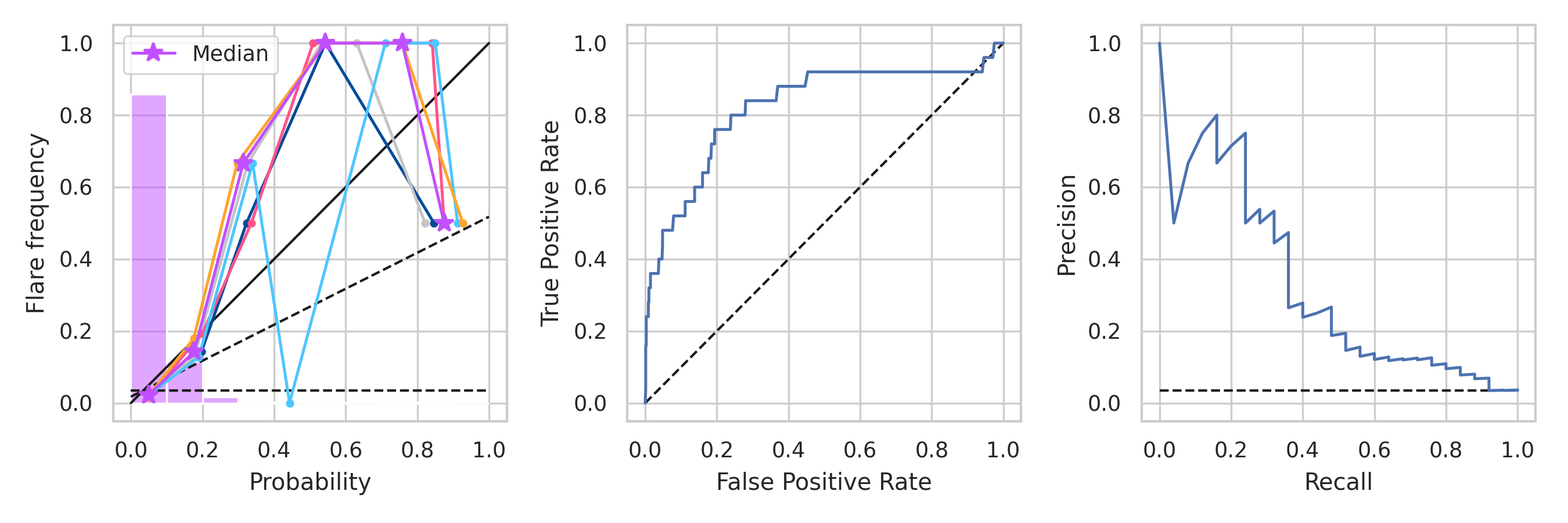}
    \includegraphics[width=\textwidth]{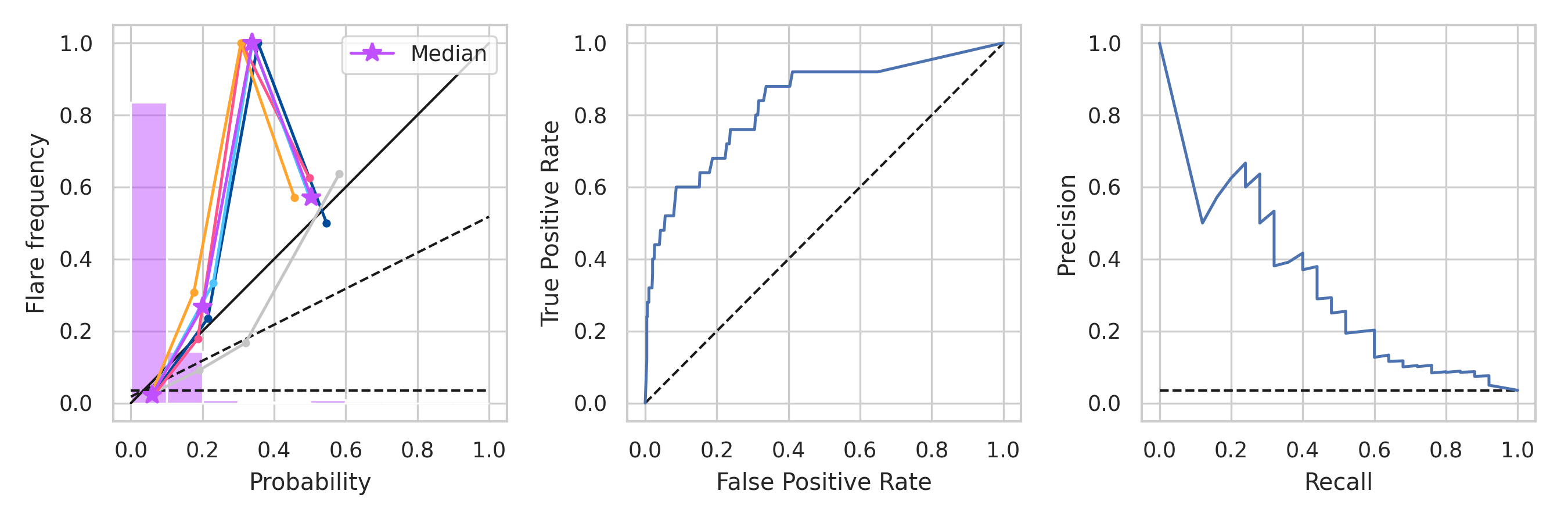}
    \caption{Performance of the calibrated CNN+LR ensemble on historical (top) and HMI (bottom) data, evaluated on Test B (2016-2017). Left, reliability diagram for the ensemble members and ensemble median. The dashed horizontal line indicates the climatological flare rate, the dashed diagonal line indicates no-skill and the solid line indicates perfect reliability. The histogram shows the distribution of output probabilities for the ensemble median. Center, the ROC curve and left, the precicison-recall curve for the calibrated ensemble median with no-skill indicated by a dashed line.}
    \label{fig:testb_performance}
\end{figure}

As expected, evaluating the models on data from 2016-2017 is a more challenging task due to how few flares occur in this period. Looking at the metrics in Table \ref{tab:testb}, neither model performs notably better than the other. The reliability diagrams shown in Fig. \ref{fig:testb_performance} are worse than on any other evaluation set and again we see that the calibrated HMI-model is not able to provide forecasts greater than 0.5 probability. Our results are in line with, but not an improvement upon, the average Brier skill scores and Gini scores of the models compared in \citet{Leka2019b}.

Due to the relative quietude of this time period compared to the solar activity over the past 40 years as depicted in Fig. \ref{fig:flareshistorical} and the climatology summary in Table \ref{tab:datasetclimatology}, we suggest that this 2016-2017 set not be used as a benchmark by the flare forecasting community moving forward. Optimizing a model for performance on this time period would likely result in decreased skill over a longer timespan. 

\begin{table}[htbp]
    \centering	
    \begin{tabular}{lccccccc}
        Training & MSE$_\downarrow$ & BSS$_\uparrow$ & APS$_\uparrow$ & Gini$_\uparrow$ & ECE$_\downarrow$ & MCE$_\downarrow$ & Max Output$_\uparrow$\\ \hline
        Historical & 0.03 (0.00) & 0.13 (0.03) & 0.32 (0.04) & 0.64 (0.06) & 0.03 (0.00) & 0.46 (0.03) & 0.87 (0.04)\\
        HMI & 0.03 (0.00) & 0.13 (0.02) & 0.30 (0.02) & 0.67 (0.02) & 0.04 (0.00) & 0.66 (0.24) & 0.51 (0.05)\\
    \end{tabular}
    \caption{Probability calibrated ensemble-median results for a 24 hour flare forecast on test B (2016-2017). The historical CNN+LR model is trained n data over all instruments from 1975-2017. The HMI CNN+LR model is trained on only SDO/HMI data from 2010-2017. {Subscript arrows indicate the direction of improvement for the given metric.}}
    \label{tab:testb}
\end{table}

\subsection{Impact of number of training years on performance}

\begin{figure}[hbtp]
    \centering
    \includegraphics[width=\textwidth]{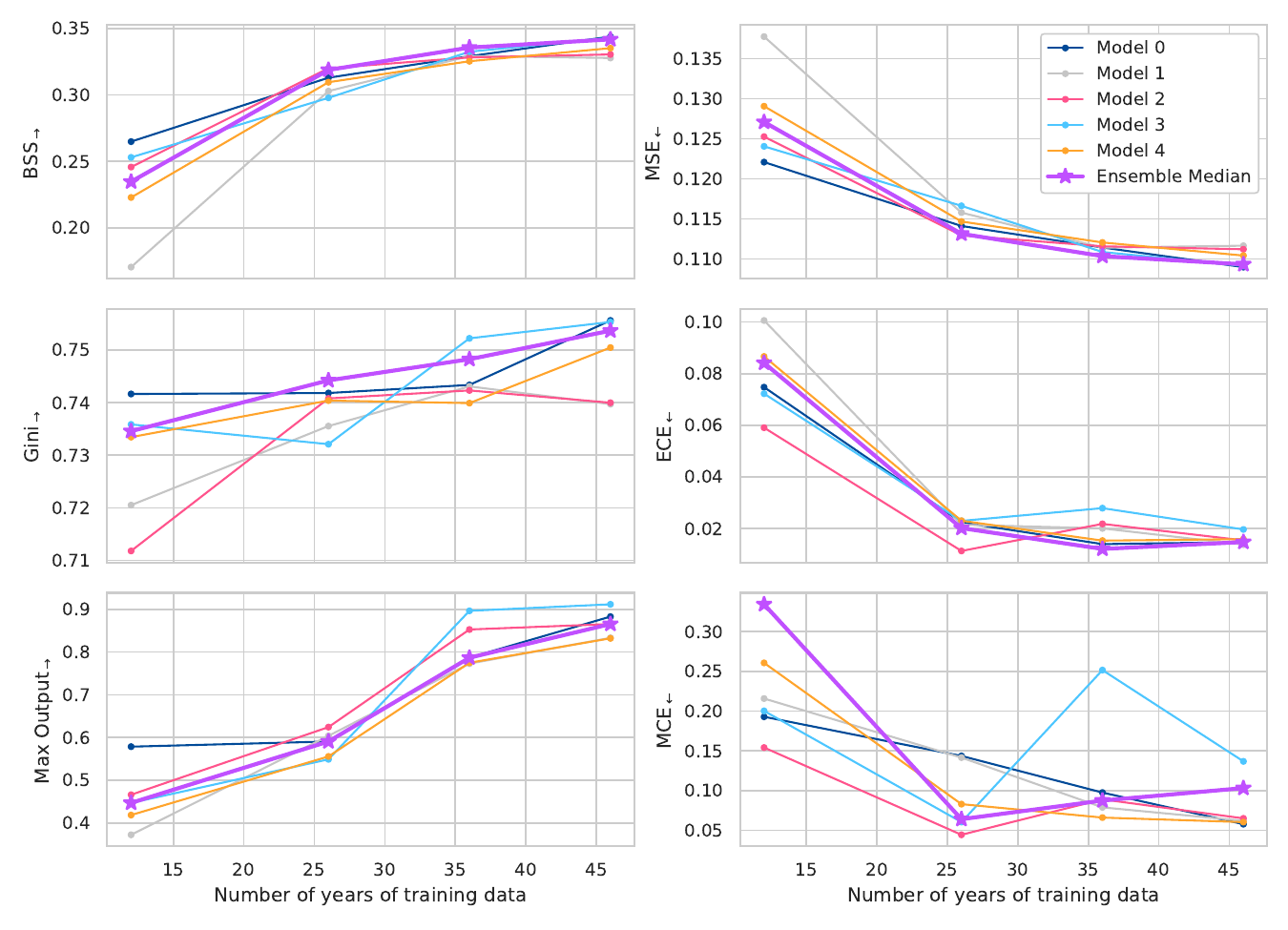}
    \caption{Performance of the calibrated CNN+LR ensemble on Test B (Nov-Dec 1976-2022) as a function of number of years of training data. Left, metrics to be maximized (BSS, Gini, Max output probability). Right, metrics to be minimized (MSE, ECE, MCE). Note that the Ensemble Median indicates the performance of the ensemble, not the median performance of the individual models.}
    \label{fig:performancevstrainingyears}
\end{figure}

As a final test of the impact of historical training data on model performance, we consider training an ensemble on varying durations of training data. We then evaluate performance as a function of number of years of training data, as seen in Fig. \ref{fig:performancevstrainingyears}. The models are tested on the same data as in Section \ref{sec:testa}, Nov-Dec 1976-2022. We can see that across all metrics other than MCE, the ensemble median performs better the more years of training data provided. In particular, the maximum probability output, which is an indication of model confidence in positive events after calibration is linearly increasing with number of years of training data. Other metrics including BSS, MSE, and ECE appear to follow a nonlinear trend, wherein performance is most improved going from 12 years to 26 years of training data. This indicates that including just two solar cycles of data for training, rather than one, is beneficial, as also observed in \citet{Sun2022}.

Both the performance of the ensemble-median and the performance of the individual ensemble members is shown in Fig. \ref{fig:performancevstrainingyears}. Here we can see one of the advantages of the ensemble approach, which is that the ensemble-median performs better across metrics than any individual model. This also makes ensembles more robust to overfitting.

\section{Conclusion}\label{sec:conclusion}

In this work, we train ML models on historical magnetogram data in order to provide probabilistic forecasts of M1 or greater flares within the next 24 hours. We consider feeding daily full-disk magnetograms from instruments spanning 1976-2022 (MWO, KPVT/SPMG, KPVT/512, SOHO/MDI and SDO/HMI) into a CNN, with scalar features computed from the magnetograms and flaring history incorporated in a final fully connected layer. We apply a probability calibration post-processing procedure to improve reliability of forecasts. Based on our experiments and comparisons with models trained on the HMI era alone, we find the following results:
\begin{itemize}
    \item Training on historical data improves forecasting skill, especially with respect to probability calibration
    \item Single frame magnetograms fed into CNNs do not add significantly more predictive power than a small set of scalar features
    \item Flare history related features offer more predictive skill than magnetogram based features
    \item Both data augmentation and pre-training for a CNN model give marginal improvements
\end{itemize}

Despite the improvement in performance from training on the full historical dataset seen in Section \ref{sec:testa}, the overall forecast skill for a 24 hour M-flare or greater forecast is not significantly improved upon compared to \citet{Leka2019b}. We can also see the drop in performance when evaluated on a test set with a significantly lower flaring rate in Section \ref{sec:testb}. 

{The majority of ML-based flare forecasting in the SDO/HMI era has focused on binary classification. Here we explicitly do not convert our forecasts to binary ones in order to emphasize the importance of moving the community towards reliable probabilistic forecasts. This means that we do not compare our results directly to the TSS metric reported in much of the literature. In addition to the reliance on binary forecasting metrics, there are not commonly used benchmark test sets for flare forecasting, which further increases the difficulty of direct comparison. For example, many works do not consider the operational setting, but artificially balance test sets and remove challenging data such as C-class flares, i.e., \citet{Chen2019,Sun2022}.} 

{To contextualize our results, we now compare to the few recent works which offer probabilistic metrics.} \citet{Nishizuka2020} report {slightly} higher BSS {(0.3)} and Gini {(0.86)} metrics for their probabilistic forecasts based on 79 input features extracted from HMI active region magnetograms, extreme ultraviolet and X-ray data for a 2015 testing period with a similar climatology to our 2016-2017 test set. {\citet{Sun2022} obtain significantly higher BSS (as high as 0.8) for their LSTM and CNN models trained on scalar features and active region magnetograms over the MDI and HMI era. However, in this study they only consider samples which exhibit either strong flares within a 24 hour forecast window ($\geq$ M1) or contain no flares of any magnitude both within the sample and forecast window. This is a much easier classification problem. They also perform balancing of both the training and test set, which further removes challenges of an operational setting. \citet{Guerra2020} generate probabilistic weighted ensembles of six "near-operational" member models available online. They report BSS ranging from 0.11 to 0.13 and Gini scores ranging from 0.54-0.77 for 24 hour M-class flare forecasts in an operational setting using data from 2014-2016. These results are well in line with our results for 2016-2017 data in Section \ref{sec:testb}.} 

{There are tradeoffs in using historical magnetograms. Although they span a longer time period, they are limited to LOS measurements and have lower spatial and temporal resolution. Many of the physics-based features used in ML-based flare forecasting on HMI data are computed from the vector magnetic field \citep{bobra:2014}. However, topological features and CNN-obtained shape-based features can be derived from LOS data alone, and these features have been shown to be similarly predictive to physics-based features \citep{Deshmukh2020,deshmukh2022decreasing,Guastavino2022,Sun2021}.} It is possible that inclusion of additional scalar features could improve performance. Although we are limited to daily LOS magnetograms in the available historical data, one could include features such as length of the strong-gradient neutral line and total magnetic dissipation, as in \citet{Song2009}, or the Schrijver R-value known to be correlated with flaring \citep{Schrijver2007}. However, one would expect a CNN to be able to automatically detect features from magnetograms, and we found our CNN model alone did not outperform a LR model trained on only flaring history features. This indicates that including a temporal component may be a more useful strategy than extracting more features from a single time frame. 

As has been previously hypothesized in \citet{deshmukh2022decreasing,Jonas2018}, it appears that we may be at the limit of magnetogram-based binary flare forecasting. Although inclusion of historical instruments allows us to capture flares over 4 solar cycles and produce more reliable flare forecasts than training on the HMI era alone, we are still a long ways from having highly accurate forecasts, as seen in ROC and precision-recall curves. Given the limits of temporal and spatial resolution of historical instruments, we cannot directly translate the approach presented here to video or timeseries data. 
In future work, we will consider using transfer learning to use our ML model as a foundation for training a new model on higher temporal cadence data from the modern era, including video sequences as in \citet{Guastavino2022} and extreme ultraviolet data as in \citet{Jonas2018,Nishizuka2020,sun2023deep}. The code for our full ML pipeline is available at \url{https://github.com/SwRI-IDEA-Lab/idea-lab-flare-forecast} {and the v0 release is archived in Zenodo \citep{flaregithub}}.

\begin{acknowledgments}
This work was funded by a NASA Space Weather O2R research grant 80NSSC22K0271 to the Southwest Research Institute. 
\end{acknowledgments}

%

\vspace{5mm}
\facilities{MWO, KPVT(SPMG and 512), SOHO(MDI), SDO(HMI)}




\appendix 

\section{Magnetogram preprocessing}
\label{appendix:preprocessing}

\begin{figure}[htbp]
    \centering
    \includegraphics[width=\textwidth]{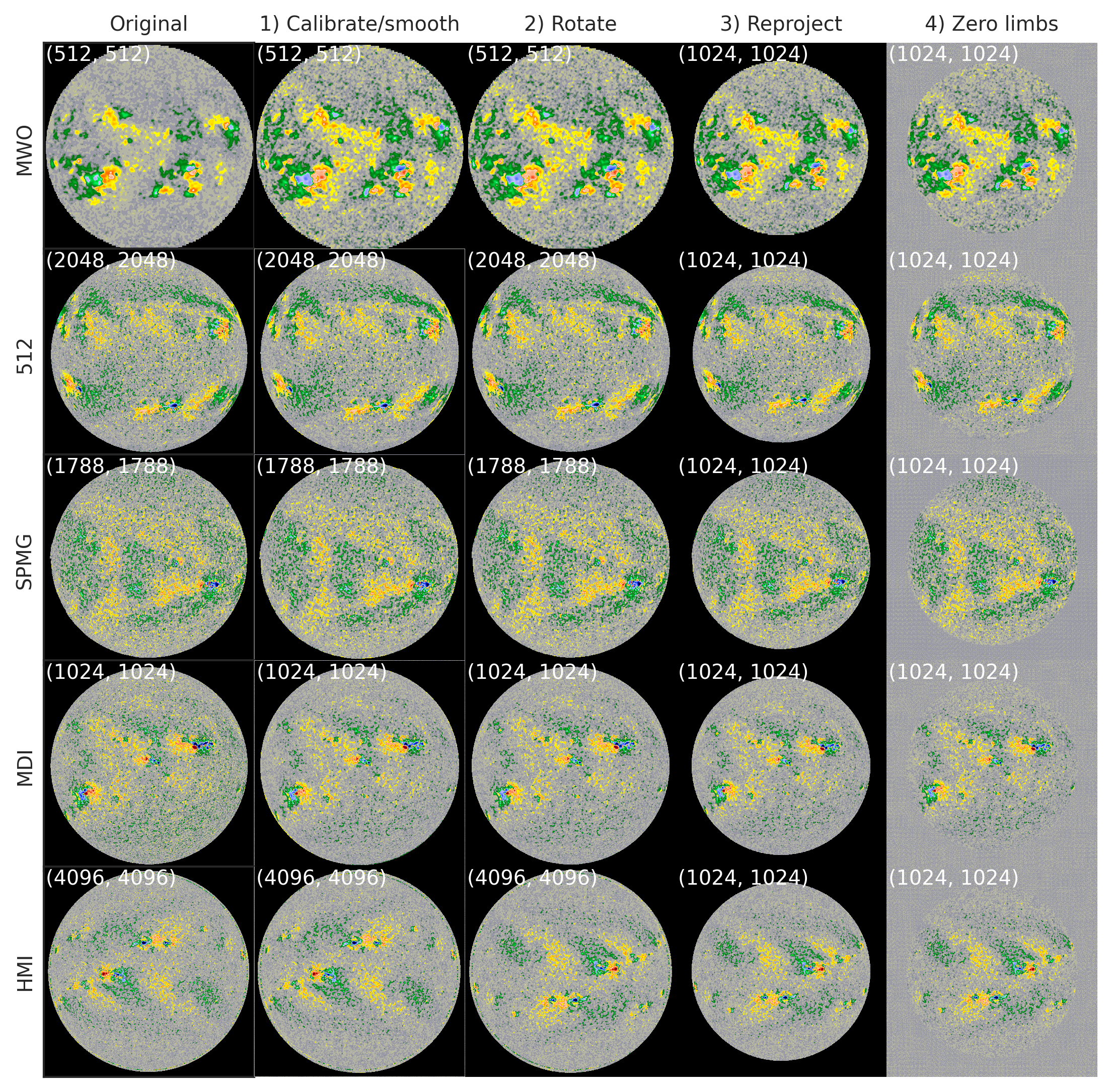}
    \caption{Preprocessing steps applied to sample magnetograms to synthesize instruments into ML-ready data. Image sizes are given in the top left corner.}
    \label{fig:preprocessing}
\end{figure}

Here we show an example of the preprocessing steps described in Section \ref{sec:preprocessing}. In Fig. \ref{fig:preprocessing} we apply preprocessing steps 1-4 to a sample magnetogram from each instrument. Most samples are already aligned with the world coordinate axis, so there is no effect from Step 2) except on the HMI sample shown.  

\section{Calibration based on total unsigned flux}
\label{appendix:totusflux}

\begin{figure}[htbp]
    \centering
    \includegraphics[width=0.48\textwidth]{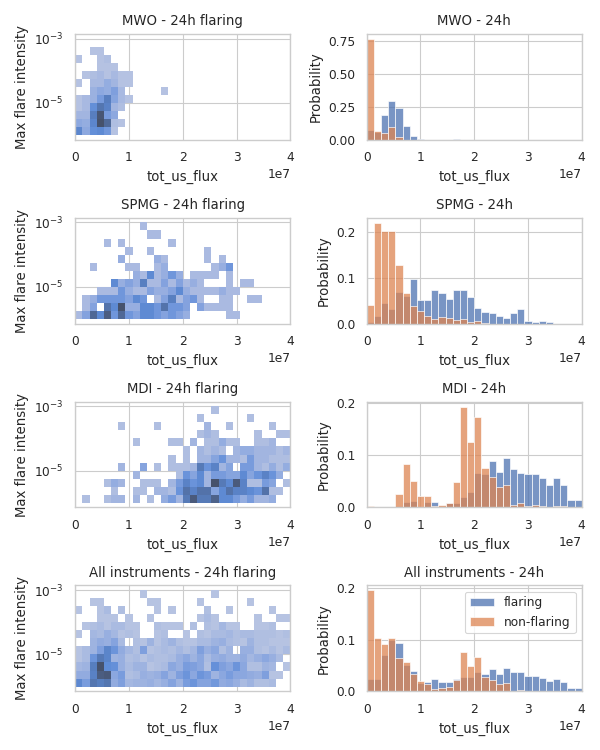}
    \includegraphics[width=0.48\textwidth]{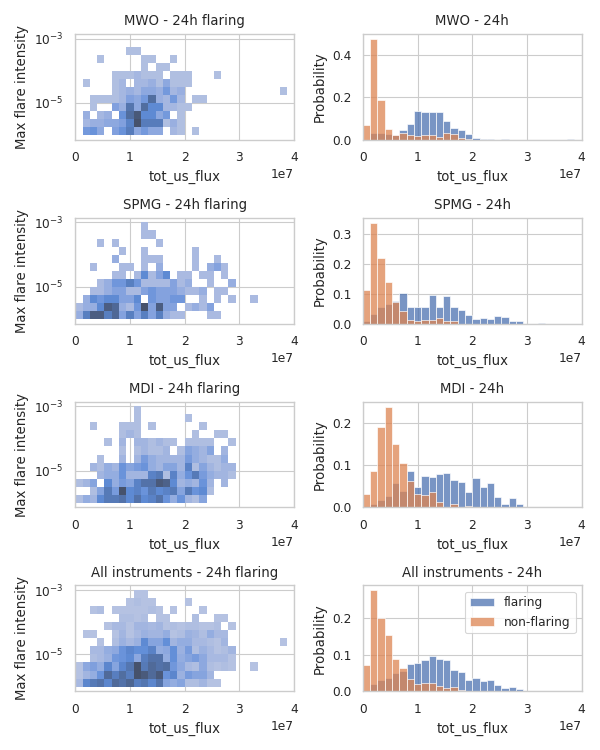}
    \caption{2-D histograms of flare intensity vs total unsigned flux on flaring samples and 1-D distributions of total unsigned flux for flaring and non-flaring samples over the period of overlap between MWO, SPMG and MDI (1996-1999). Left, before smoothing and an MWO calibration factor of 2, right, after smoothing/calibration.}
    \label{fig:flarevstotusflux_overlap}
\end{figure}

\begin{figure}[htbp]
    \centering
    \includegraphics[width=0.5\textwidth]{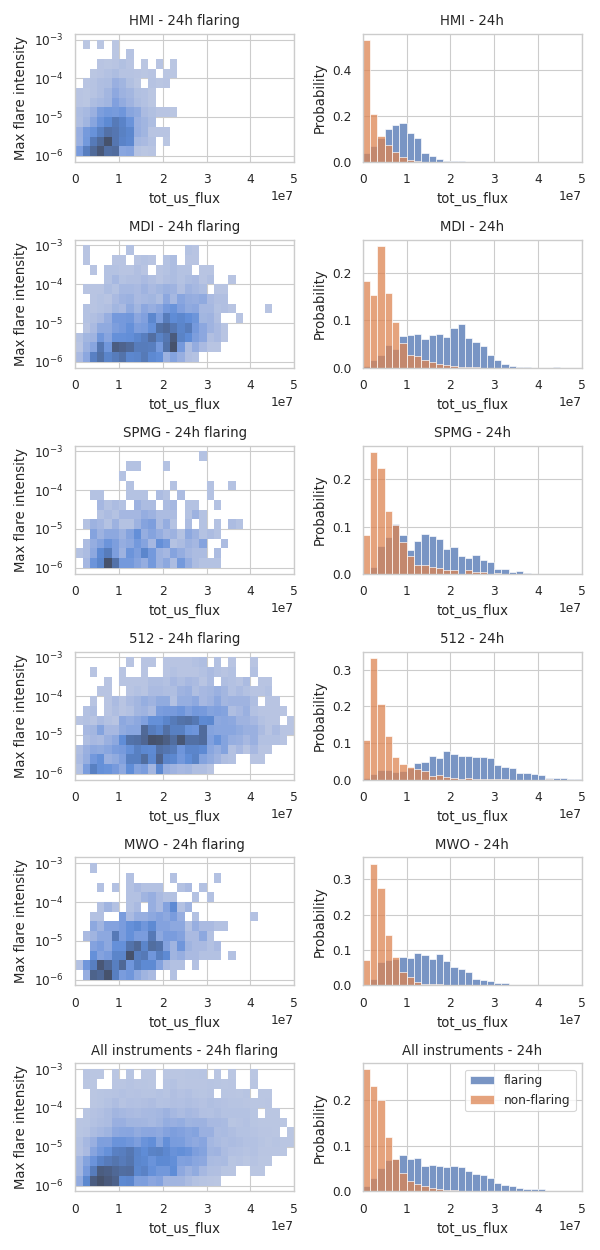}
    \caption{2-D histograms of flare intensity vs total unsigned flux on flaring samples and 1-D distributions of total unsigned flux for flaring and non-flaring samples over the dataset.}
    \label{fig:flarevstotusflux}
\end{figure}

Given the correlation of total unsigned flux with active region activity, we consider using a threshold based on total unsigned flux as a pretraining label in Section \ref{sec:pretraining}, under the assumption that this is easier to learn. In order to see how well the instruments are calibrated, determine the MWO calibration factor of 2, and to select a threshold for pretraining using unsigned flux, we investigate the 2-D distribution of total unsigned flux vs. maximum flare intensity within a 24 hour forecast window. We also look at 1-D distributions of total unsigned flux for both flaring ($\geq$ M-flare) and non-flaring samples within the same time window. Fig. \ref{fig:flarevstotusflux_overlap} shows these distributions for the period of overlap between MWO, SPMG and MDI both before and after smoothing/calibration. It is clear that applying the Gaussian smoothing to instruments other than MWO and the factor of 2 calibration to MWO improves the alignment between distributions across instruments. Fig. \ref{fig:flarevstotusflux} shows the distributions for each instrument over the final dataset with all preprocessing applied.

\bibliography{references}{}
\bibliographystyle{aasjournal}



\end{document}